\documentclass[11pt]{article}

\usepackage{epsf}
\usepackage{amsmath}
\usepackage{amssymb}
\usepackage{graphicx}
\usepackage{dcolumn}
\usepackage{bm}

\begin{document}

\def\thefootnote{\fnsymbol{footnote}}

\vspace*{-1.5cm}
\begin{flushright}
\tt MTA-PHYS-0604
\end{flushright}

\vspace{0.2cm}
\begin{center}
\Large\bf\boldmath
Constraints on the mSUGRA parameter space from NLO calculation of isospin asymmetry in $B\to K^*\gamma$
\unboldmath
\end{center}

\vspace{0.4cm}
\begin{center}
M. R. Ahmady\footnote{Electronic address: {\tt mahmady@mta.ca}; URL: {\tt http://www.mta.ca/$^\sim$mahmady}} and F. Mahmoudi\footnote{Electronic address (corresponding author): \tt nmahmoudi@mta.ca}\\[0.4cm]
{\sl Department of Physics, Mount Allison University, 67 York Street, Sackville,\\
 New Brunswick, Canada E4L 1E6}
\end{center}
\vspace{0.5cm}

\begin{abstract}
\noindent The contributions of supersymmetric particles in the isospin symmetry violation in $B\to K^*\gamma$ decay mode are investigated. The model parameters are adopted from minimal Supergravity with minimal flavor violation. A complete scan of the mSUGRA parameter space has been performed, using the next to leading supersymmetric contributions to the relevant Wilson coefficients. The results are compared to recent experimental data in order to obtain constraints on the parameter space. We point out that isospin asymmetry can prove to be an interesting observable and imposes severe restrictions on the allowed parameter space, in particular for large values of $\tan\beta$. The constraints obtained with isospin asymmetry also appear as more restricting than the ones from the branching ratio of $B\to X_s\gamma$.
\end{abstract}
\vspace{0.3cm}
\section{Introduction}
\noindent The Standard Model (SM) has been very successful to describe the experimental data from accelerator physics so far. With new colliders, like the Large Hadron Collider (LHC) or later on the International Linear Collider (ILC), becoming operational, the hope is to detect signals which could reveal physics beyond the SM, that in turn provide answers to the many theoretical questions left unanswered by the SM.\\
\\
One of the most motivated scenarios for new physics is generally considered to be Supersymmetry (SUSY). In the minimal supersymmetric extension of the Standard Model (MSSM), the large number of free parameters makes the phenomenological studies rather complicated. Many studies are therefore based on the constrained minimal supersymmetric Standard Model (CMSSM) -- often called mSUGRA -- with the number of parameters reduced to five, corresponding to $m_0$ (common mass of scalar particles at the supersymmetric grand unification scale), $m_{1/2}$ (universal gaugino mass), $A_0$ (universal trilinear SUSY breaking parameter), together with the sign of the Higgs mixing parameter $\mu$ and the ratio of the two Higgs vacuum expectation values $\tan\beta$.\\
\\
Many studies have been performed to constrain the supersymmetric parameter space, and in particular direct and indirect searches for new particles have provided lower bounds on their masses \cite{PDG2006}. Other constraints come from the cosmological observations of the large scale structures and the cosmic microwave background \cite{wmap}, the measurement of the anomalous magnetic moment of the muon ($g_\mu -2$) \cite{g_2}, and the study of radiative B meson decays.\\
\\
The precision measurements of the radiative B meson decays, which have become possible with the operation of the B factories and other B-dedicated experiments, have provided exciting opportunities for mapping possible routes beyond the SM. One such rare decay mode is the exclusive process $B\to K^*\gamma$ and its
associated inclusive transition $b\to s\gamma$, which have been extensively used to constrain new physics \cite{battaglia,allanach}. 
Consequently, a thorough investigation of the branching ratio of these decay modes has been instrumental in constraining the parameter space of various models. 
In this paper, we focus on another observable, the isospin asymmetry, and we will show that this observable may even lead to more stringent constraints than the branching ratios.\\
\\
The isospin asymmetry for the exclusive process $B\to K^*\gamma$ is defined as:
\begin{equation}
\Delta_{0-}=\frac{\Gamma (\bar B^0\to\bar K^{*0}\gamma ) -\Gamma (B^-\to K^{*-}\gamma )}{\Gamma (\bar B^0\to\bar K^{*0}\gamma )+\Gamma (B^-\to K^{*-}\gamma
)}\;\; , \label{isospinasym}
\end{equation}
with $\Delta_{0+}$ obtained from eq.(\ref{isospinasym}) by using the charge conjugate modes. 
The most recent data for exclusive decays from Belle \cite{belle} and Babar \cite{babar} point to isospin asymmetries of at most a few percent, consistent with zero within the experimental errors:
\begin{eqnarray}
\Delta_{0-}&=& +0.050 \pm 0.045({\rm stat.})\pm 0.028({\rm syst.})\pm 0.024(R^{+/0})\;\;\; (\mbox{Babar})\; , \label{babar}\\
\Delta_{0+}&=& +0.012 \pm 0.044({\rm stat.})\pm 0.026({\rm syst.})\;\;\; (\mbox{Belle})\; , \label{belle}
\end{eqnarray}
where the last error in eq.(\ref{babar}) is due to the uncertainty in the ratio of the branching fractions of the neutral and charged B meson production in $\Upsilon (4S)$ decays. Within the SM, this asymmetry, which is due to the non-spectator contributions, has been estimated in the literature, using the QCD factorization approach in Refs. \cite{kagan} and \cite{bosch},  Brodsky-Lepage formalism \cite{petrov}, and the perturbative QCD method \cite{keum}. On the other hand, Ref. \cite{ac} deals with the effects of an additional generation of vector-like quarks on the isospin symmetry breaking in $B\to K^*\gamma$. We consider here the QCD factorization method for our calculations.\\
\\
In the following, we first present the general framework of our investigation, followed by an analysis of the constraints of the isospin asymmetry on the mSUGRA parameter space in the context of minimal flavor violation. In order to compare the constraints due to isospin breaking with those obtained from the branching ratios, the results for inclusive branching ratio are reproduced as well.
\section{General framework}
\noindent The effective Hamiltonian for $b\to s\gamma$ transitions reads:
\begin{equation}
{\cal H}_{eff}=\frac{G_F}{\sqrt{2}}\sum_{p=u,c} V^*_{ps}V_{pb} \left[ C_1(\mu) \, O^p_1(\mu) + C_2(\mu) \, O^p_2(\mu) +\sum_{i=3}^8 C_i(\mu) \, O_i(\mu)\right]\;\;,
\end{equation}
where $G_F$ is the Fermi coupling constant, $V_{ij}$ are elements of the CKM matrix, $O_i(\mu)$ are the operators relevant to $B\to K^*\gamma$ and $C_i(\mu)$ are the corresponding Wilson coefficients evaluated at the scale $\mu$. Since the combination $V^*_{us}V_{ub}$ is an order of magnitude smaller than $V^*_{cs}V_{cb}$, we can safely neglect the u-quark terms. The operators $O_i$ can be listed as follows:
\begin{eqnarray}
\nonumber O_1^p&=&\bar s_\alpha\gamma^\mu P_Lp_\beta\bar p_\beta\gamma_\mu P_Lb_\alpha \;\; ,\raisebox{0cm}[0cm][0.4cm]{~}\\ 
\nonumber O_2^p&=&\bar s_\alpha\gamma^\mu P_L p_\alpha\bar p_\beta\gamma_\mu P_Lb_\beta \;\; ,\raisebox{0cm}[0cm][0.3cm]{~}\\
\nonumber O_3&=&\bar s_\alpha\gamma^\mu P_Lb_\alpha\sum_{q}\bar q_\beta\gamma_\mu P_Lq_\beta \;\; ,\\
O_4&=&\bar s_\alpha\gamma^\mu P_Lb_\beta\sum_{q}\bar q_\beta\gamma_\mu P_Lq_\alpha \;\; ,\\
\nonumber O_5&=&\bar s_\alpha\gamma^\mu P_Lb_\alpha\sum_{q}\bar q_\beta\gamma_\mu P_Rq_\beta \;\; ,\\
\nonumber O_6&=&\bar s_\alpha\gamma^\mu P_Lb_\beta\sum_{q}\bar q_\beta\gamma_\mu P_Rq_\alpha \;\; ,\\
\nonumber O_7 &=&\frac{e}{4\pi^2} m_b \bar s_\alpha\sigma^{\mu\nu} P_R b_\alpha F_{\mu\nu} \raisebox{0cm}[0cm][0.4cm]{~}\;\; ,\\
\nonumber O_8 &=&\frac{g_s}{4\pi^2} m_b \bar s_\alpha\sigma^{\mu\nu} P_R T^a_{\alpha\beta}b_\beta G_{\mu\nu}^a \raisebox{0cm}[0cm][0.2cm]{~}\;\;,
\label{operators}
\end{eqnarray}
where $P_L(P_R)=\dfrac{1-(+)\gamma_5}{2}$ are the projection operators. The electroweak penguin operators are omitted from the above list as their contributions to the process at hand are negligibly small compared to the others. \\
\\
The presence of SUSY particles does not introduce new operators in the list, however, the Wilson coefficients $C_i$ receive additional contributions from virtual sparticles.\\
We use the expressions of the Wilson coefficients at the next to leading order (NLO) in the strong coupling constant $\alpha_s$. They are first calculated at the scale $\mu_W = O(M_W)$. The contributions from the W boson ($SM$), the charged Higgs ($H$) and the charginos ($\chi$), as well as the leading $\tan\beta$ corrections to the W boson and the charged Higgs are considered:
\begin{eqnarray}
C_i(\mu_W)=C_i^{SM}(\mu_W)+\delta C_i^{H}(\mu_W)+\delta C_i^{\chi}(\mu_W)+\delta C_i^{(SM,\tan\beta)}(\mu_W)+\delta C_i^{(H,\tan\beta)}(\mu_W)\;\;.
\end{eqnarray}
The contributions from the gluino and neutralino are neglected in our work, as they are known to be negligible in the minimal flavor violating scenario~\cite{bertolini}. 
The reason is that within mSUGRA framework, there exists a strong correlation among various parameters. In particular, the down squarks, which appear in the gluino and neutralino loops are much heavier than the stops, the virtual partner in the chargino loop. In fact, chargino can be relatively light and since at least one of the stops can also be light, their loop results in a considerable contribution.\\
\\
The details of the calculation of the Wilson coefficients at the scale $\mu_W$ are given in Appendix A.  
The Wilson coefficients are then evolved through the renormalization group equations to the scale $\mu_b = O(m_b)$, at which they can be used to calculate the isospin asymmetry. Further details are given in Appendix B.\\
\\
Following the method of Ref. \cite{kagan}, one can write the nonspectator isospin symmetry breaking contribution as $A_q=b_qA_{lead}$, where $q$ is the flavor of the light anti-quark in the B meson and $A_{lead}$ is the leading isospin symmetry conserving spectator amplitude. To leading order in $\alpha_s$, the main contribution to $B\to K^*\gamma$ is from the electromagnetic penguin operator $O_7$: 
\begin{equation}
A_{lead} = -i\frac{G_F}{\sqrt{2}}V_{cb}V_{cs}^*\,a_7^c\langle K^*\gamma|O_7|B\rangle\;\; .\label{Alead}
\end{equation}
The factorizable amplitude $A_{lead}$ is proportional to the form factor $T_1^{B\to K^*}$ which parameterizes the hadronic matrix element of $O_7$ to the leading order in $\Lambda_{QCD}/m_b$. The coefficient $a_7^c$, which is dominated by $C_7$, is defined in Appendix C. $b_q$ depends on the flavor of the spectator and, in fact, the above parameterization leads to a simple expression for the isospin asymmetry in terms of this parameter:
\begin{equation}
\Delta_{0-}=\mbox{Re} (b_d-b_u)\;\; .\label{asymb}
\end{equation}
The expression for $b_q$, which is derived in \cite{kagan} within the QCD factorization method, can be found in Appendix C.\\
\\
In order to generate the SUSY mass spectrum, as well as the couplings and the mixing matrices, we use the Monte Carlo event generator ISAJET-7.74 \cite{baer}. We perform scans in the mSUGRA parameter space ($m_0$, $m_{1/2}$, $A_0$, $\mbox{sign}(\mu)$, $\tan\beta$). For any mSUGRA parameter space point, we then calculate the isospin asymmetry using eq.(\ref{asymb}), and compare it to the combined experimental limits of eq.(\ref{babar}) and eq.(\ref{belle}). After including the theoretical errors due to the scales and model parameters, we allow mSUGRA parameter space points which stand in the $95\%$ confidence level range
\begin{equation}
-0.047 < \Delta_{0-} < 0.093 \;\;.\label{isospinconstraint}
\end{equation}
For comparison, we also perform the calculation of the inclusive branching ratio of $B\to X_s\gamma$ following Ref. \cite{kagan99}, and allow the mSUGRA parameter space points to be in the $95\%$ confidence level range \cite{battaglia}
\begin{equation}
2.33 \times 10^{-4} < \mathcal{B}(B\to X_s\gamma) < 4.15 \times 10^{-4} \;\;.\label{BRconstraint}
\end{equation}
The points which result in too small light Higgs masses ({\it i.e.} such as $m_{h^0} < 111$ GeV) or which do not satisfy the constraints presented in Table \ref{bounds} are also excluded.\\
\begin{table}[!ht]
\begin{center}\renewcommand{\arraystretch}{1.5}
\begin{tabular}{|c|c|c|c|c|c|c|c|c|c|} \hline
Particle & ~~$\chi^0_1$~~ & ~~$\tilde{l}_R$~~ & ~$\tilde{\nu}_{e,\mu}$~ & ~~$\chi^\pm_1$~ & ~~$\tilde{t}_1$~~ & ~~$\tilde{g}$~~~ & ~~$\tilde{b}_1$~~ & ~~$\tilde{\tau}_1$~~ & ~~$\tilde{q}_R$~~\\
\hline
Lower bound & 46 & 88 & 43.7 & 67.7 & 92.6 & 195 & 89 & 81.9 & 250\\
\hline
\end{tabular}
\caption{Lower bounds on sparticle masses in GeV, obtained from \cite{PDG2006}.}
\label{bounds}
\end{center}
\end{table}%
\\
Finally, we also examine whether the lightest supersymmetric particle (LSP) is charged. Indeed, 
the LSP is stable when R-parity is conserved, and to be accounted for dark matter, it has to be neutral. On the other hand, if R-parity is violated, then the LSP is not stable and as such cannot be a candidate for the dark matter. In this case, it is possible to have charged LSP with no constraint from cosmology. In our results, we have identified the parameter space regions where the LSP is charged to indicate the cosmologically disfavored mSUGRA parameters if R-parity is conserved.\\
\\
Our results for $B\to X_s\gamma$ inclusive branching ratio are consistent with those from Ref. \cite{battaglia} and from the MicrOMEGAs code \cite{micromegas}.\\
\\
An analysis of our results is presented in the following section.
\section{Constraints from isospin asymmetry}
%
\begin{figure}[!ht]
\hspace*{-0.5cm}\includegraphics[width=9cm,height=8cm]{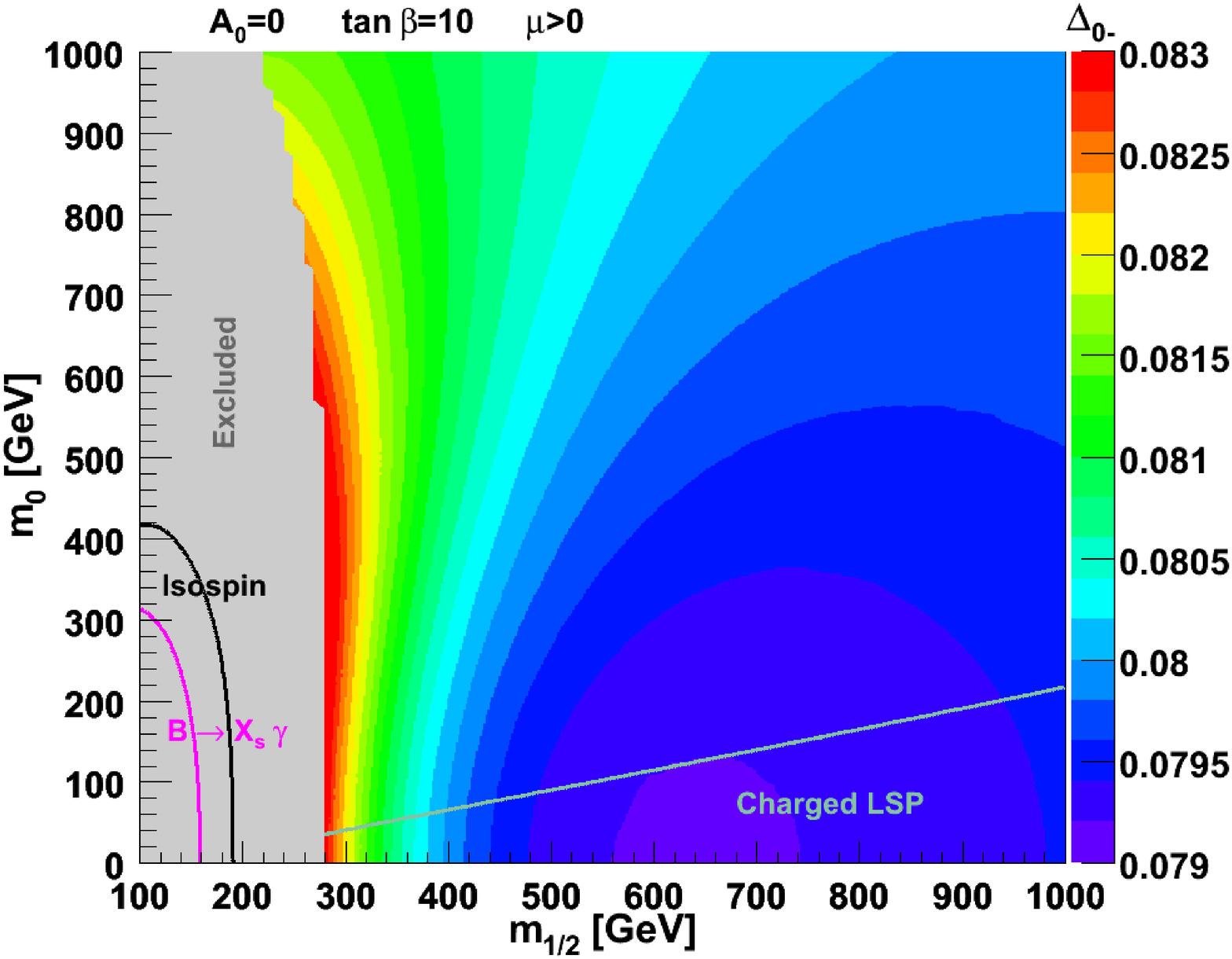}~\includegraphics[width=9cm,height=8cm]{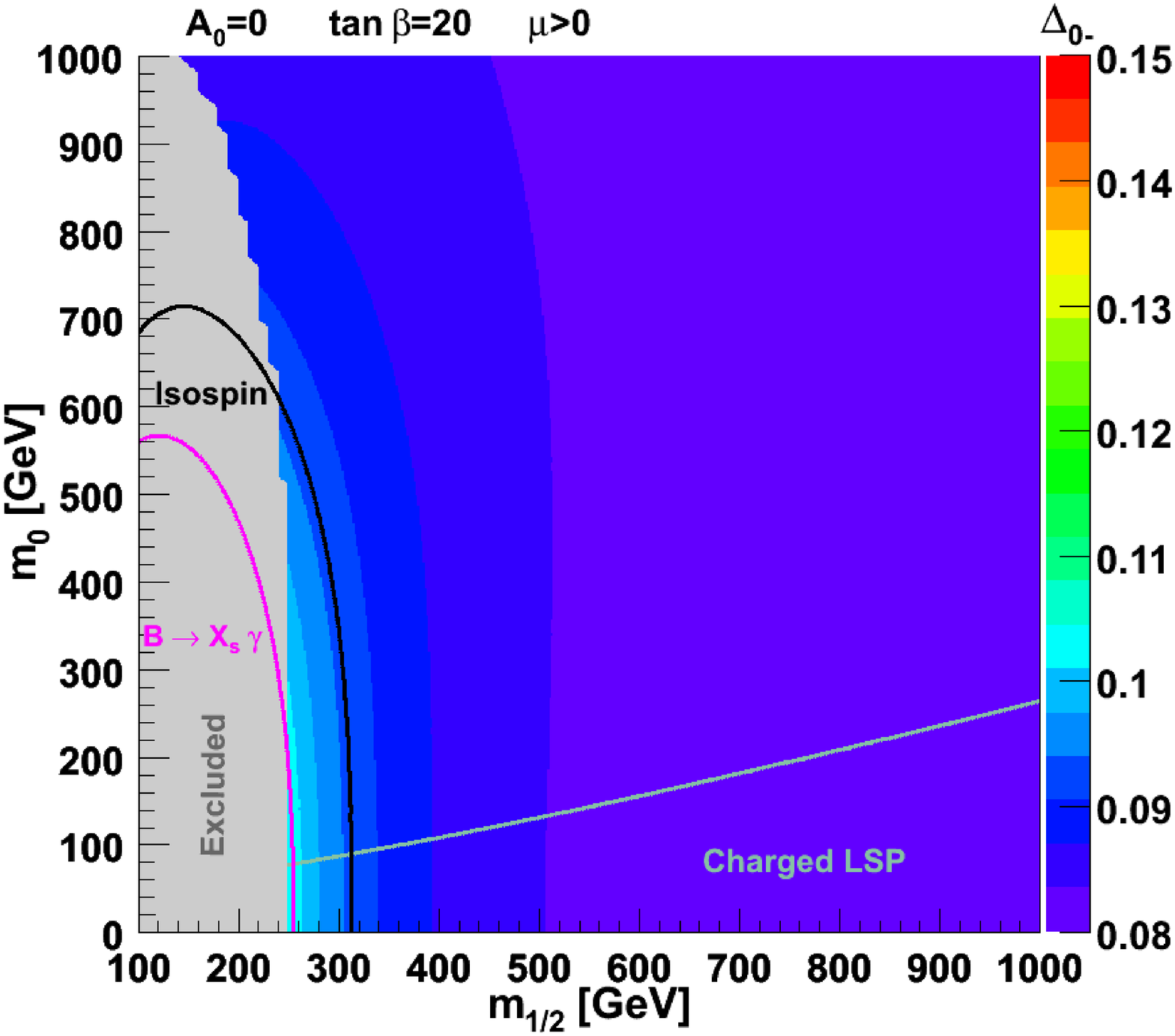}\\
\hspace*{-0.5cm}\includegraphics[width=9cm,height=8cm]{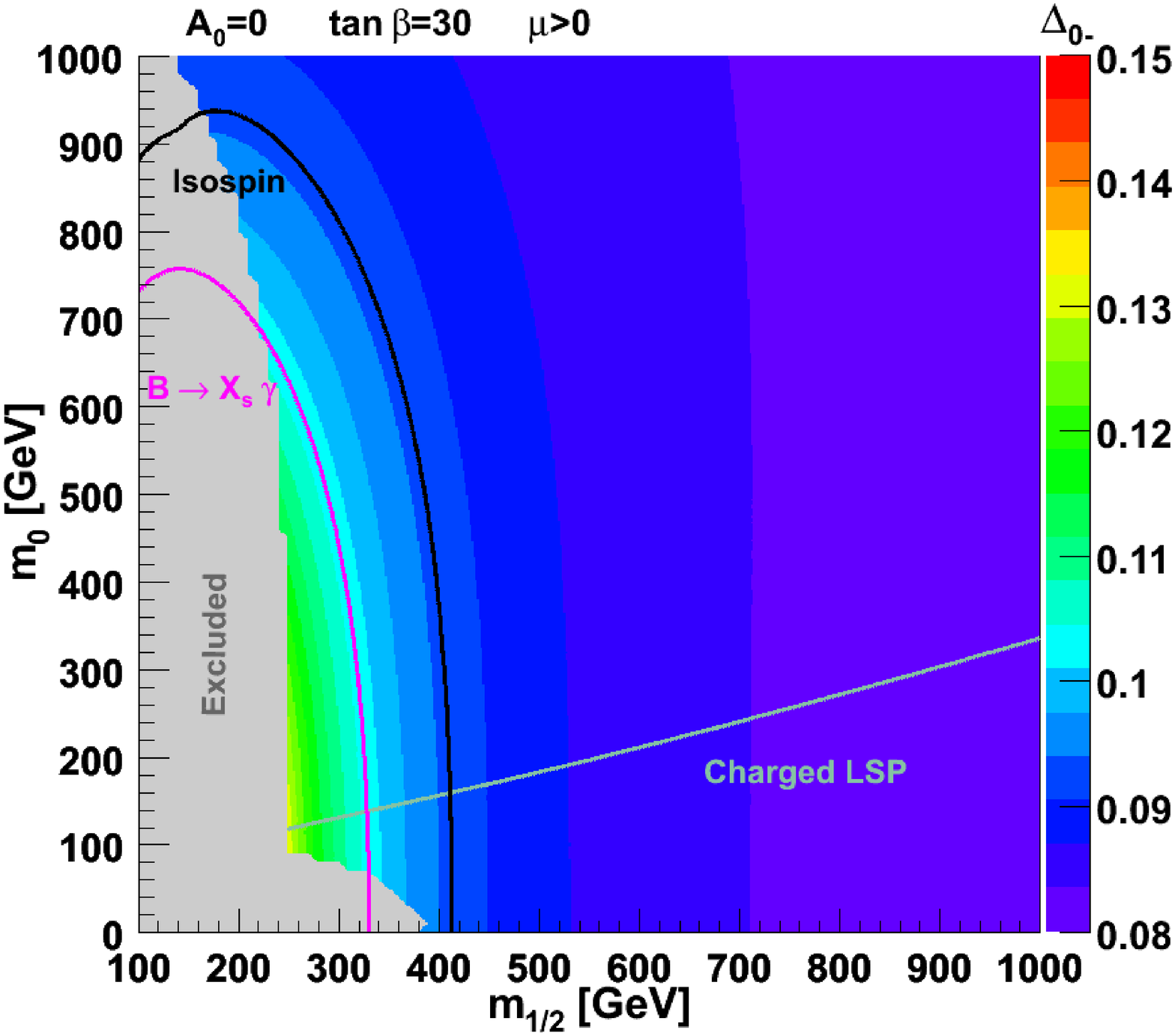}~\includegraphics[width=9cm,height=8cm]{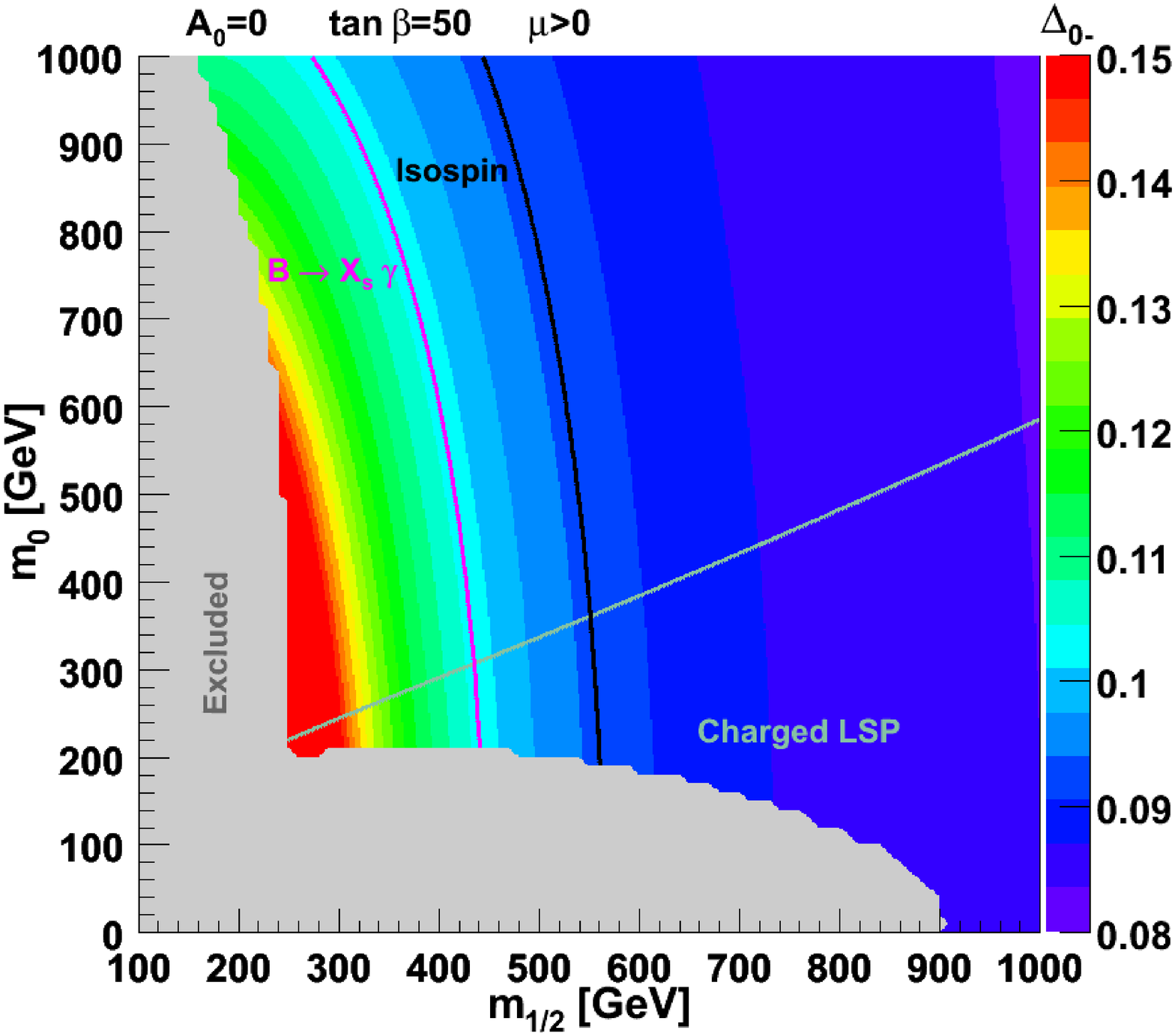}
\caption{Constraints on the mSUGRA parameter plane $(m_{1/2},m_0)$ for $A_0=0$ and for different values of $\tan\beta$. The ``Excluded'' region in gray corresponds to the sparticle or Higgs search constraints of Table~\ref{bounds}. The light green ``Charged LSP'' corresponds to a cosmologically disfavored region. The magenta ``$B\to X_s\gamma$'' contour delimits the region excluded by the inclusive branching ratio in accordance with eq.(\ref{BRconstraint}), whereas the yellow ``Isospin'' region corresponds to the isospin symmetry breaking constraints from eq.(\ref{isospinconstraint}). Note that the color scale is different for the first graph with $\tan\beta=10$.}\label{A00}
\end{figure}%
\begin{figure}[!ht]
\hspace*{-0.5cm}\includegraphics[width=9cm,height=8cm]{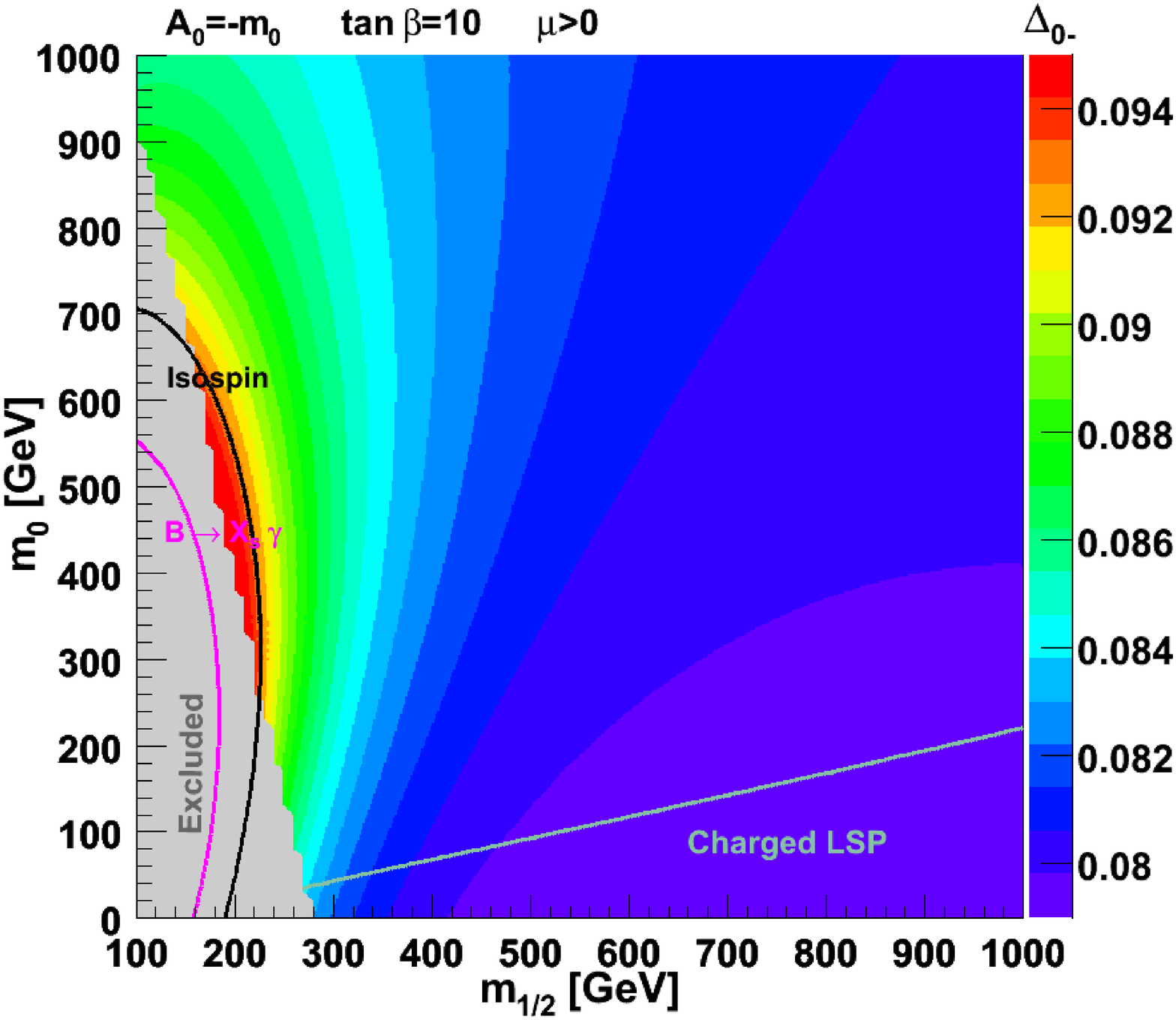}~\includegraphics[width=9cm,height=8cm]{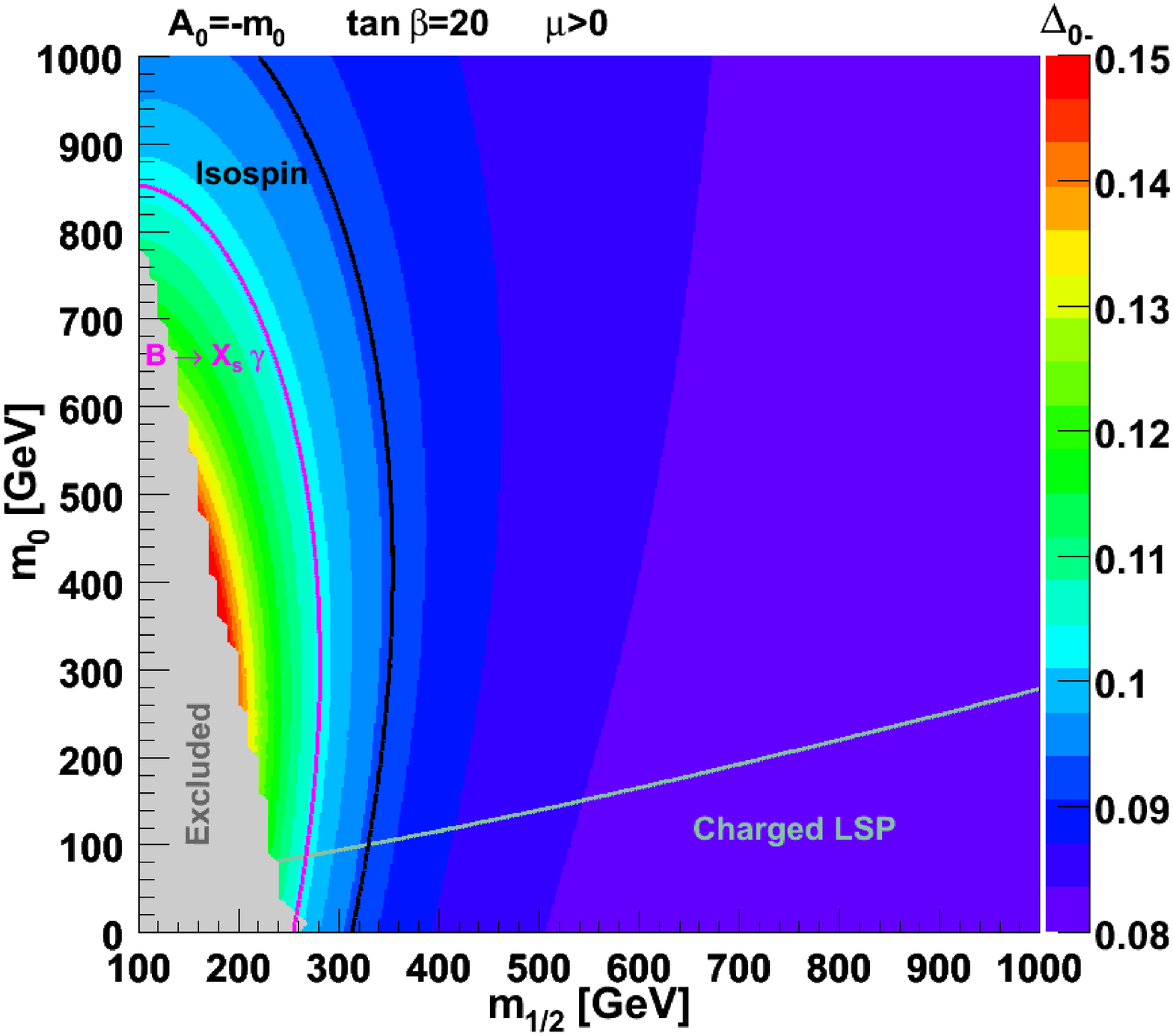}\\
\hspace*{-0.5cm}\includegraphics[width=9cm,height=8cm]{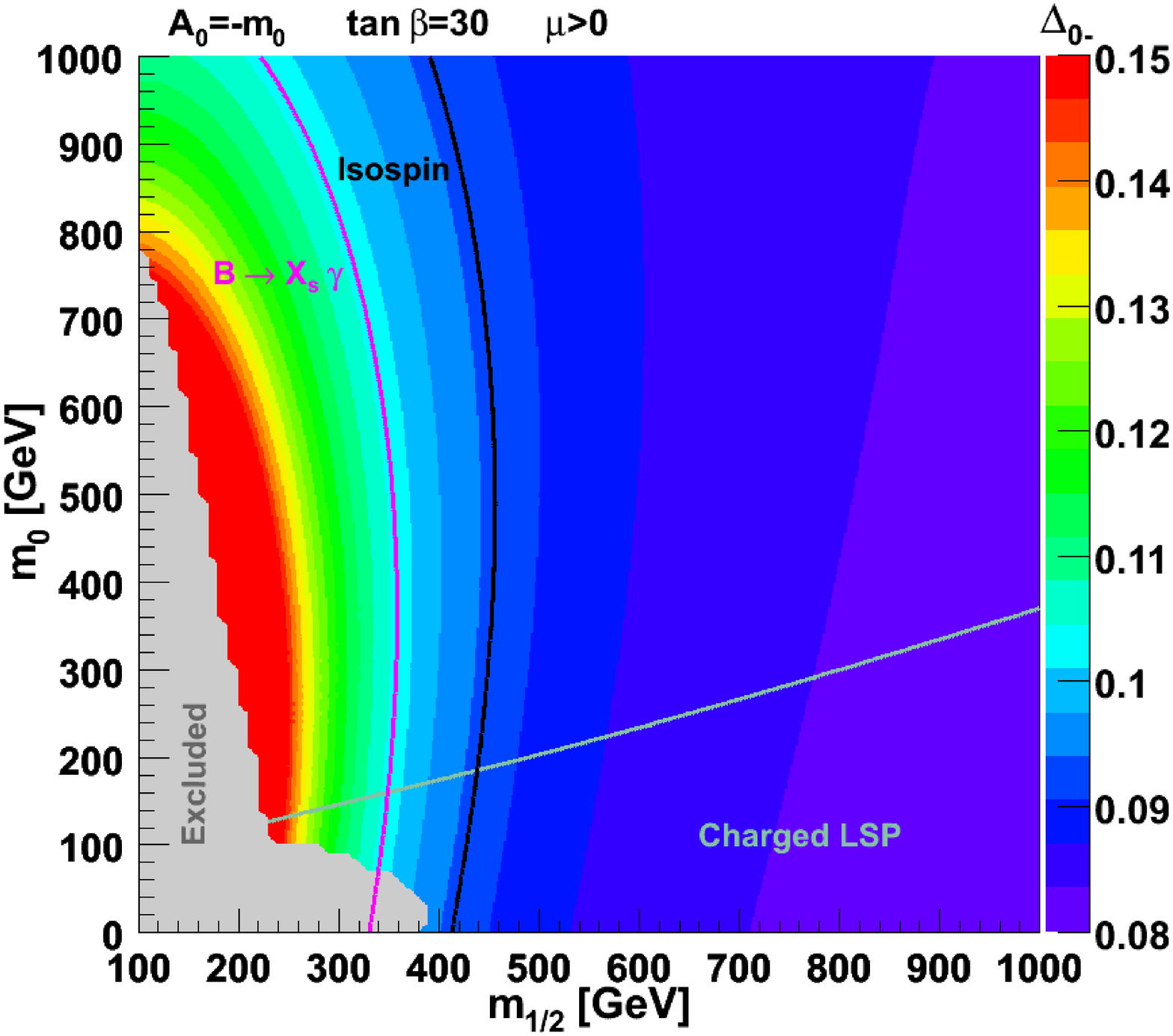}~\includegraphics[width=9cm,height=8cm]{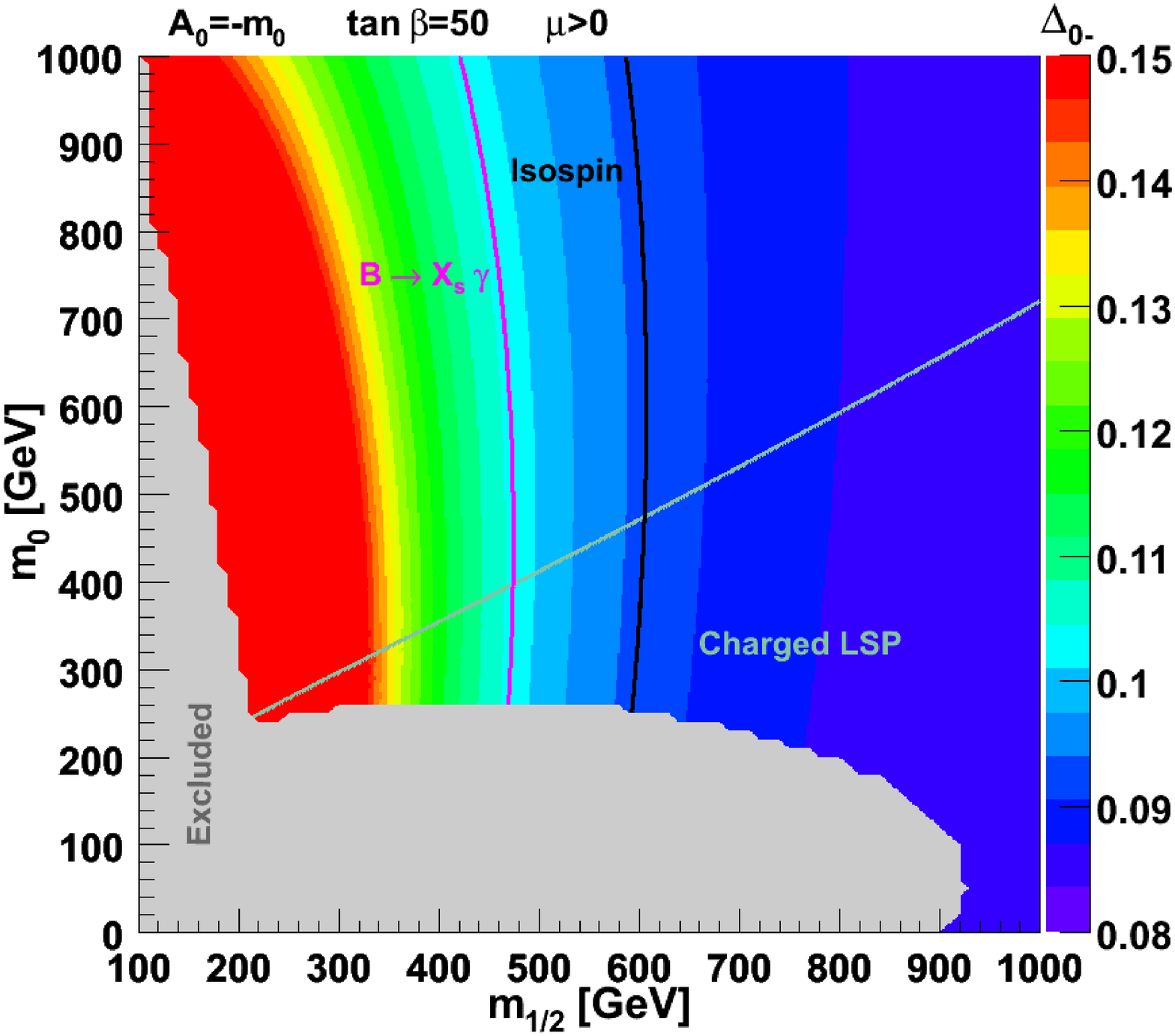}
\caption{Constraints on the mSUGRA parameter plane $(m_{1/2},m_0)$ for $A_0=-m_0$. The conventions for the different regions are the same as in Fig.~\ref{A00}.}\label{A0mm0}
\end{figure}%
\begin{figure}[!ht]
\hspace*{-0.5cm}\includegraphics[width=9cm,height=8cm]{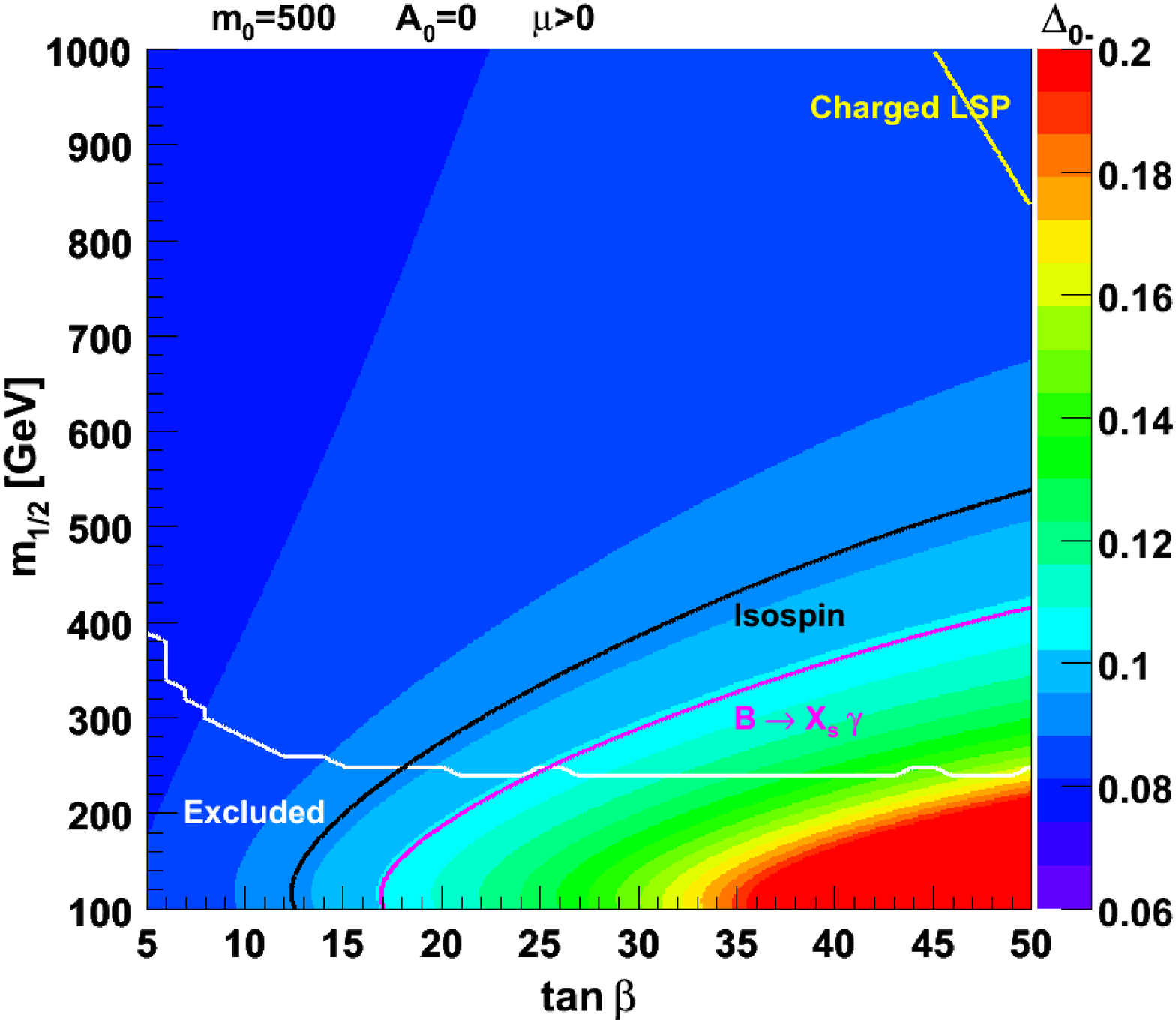}~\includegraphics[width=9cm,height=8cm]{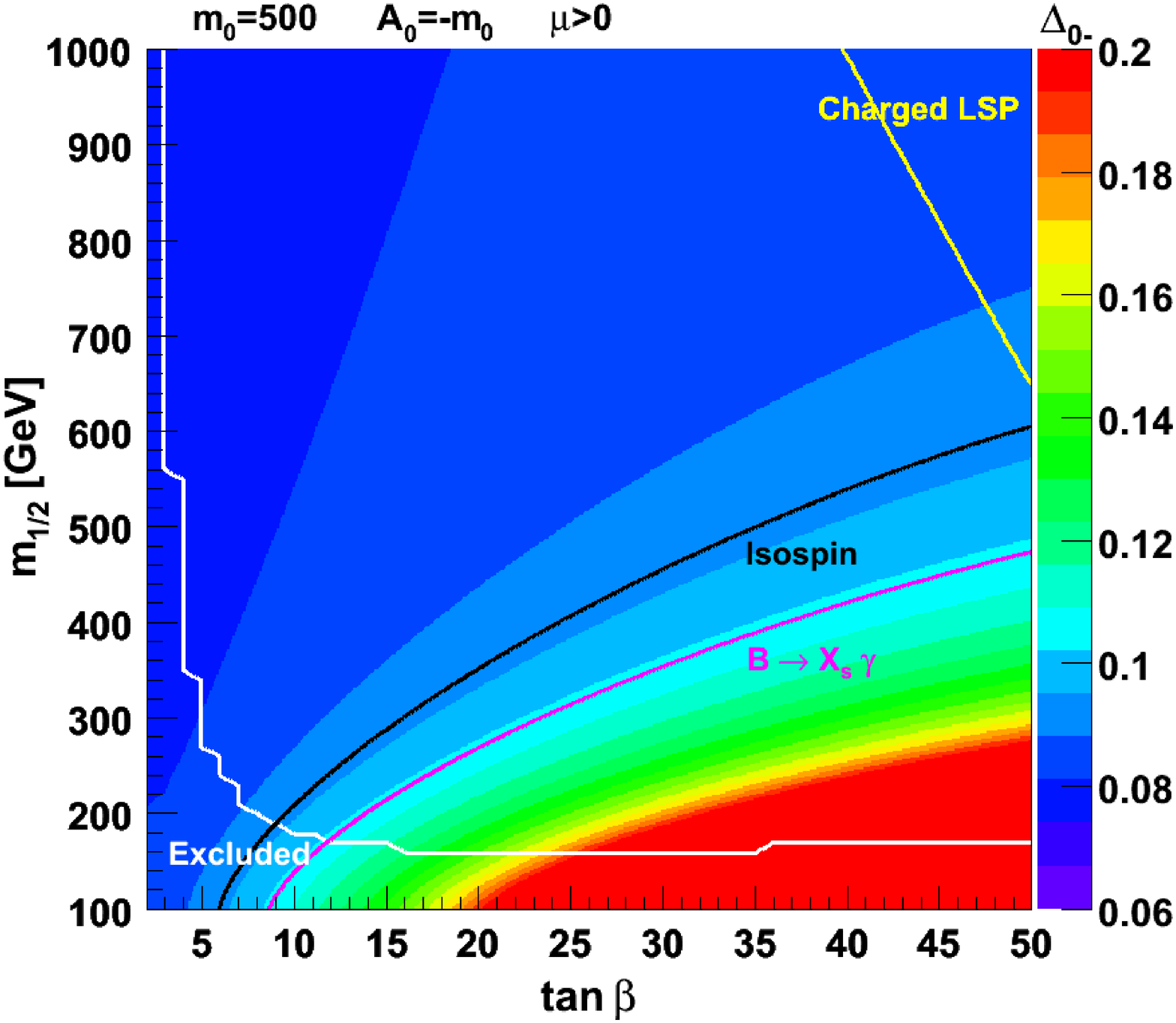}
\caption{Constraints on the mSUGRA parameter plane $(\tan\beta,m_{1/2})$ for $m_0=500$, with $A_0=0$ and $A_0=-m_0$. The definitions of the different regions are in the text.}\label{tanb}
\end{figure}%
\begin{figure}[!ht]
\hspace*{-0.5cm}\includegraphics[width=9cm,height=8cm]{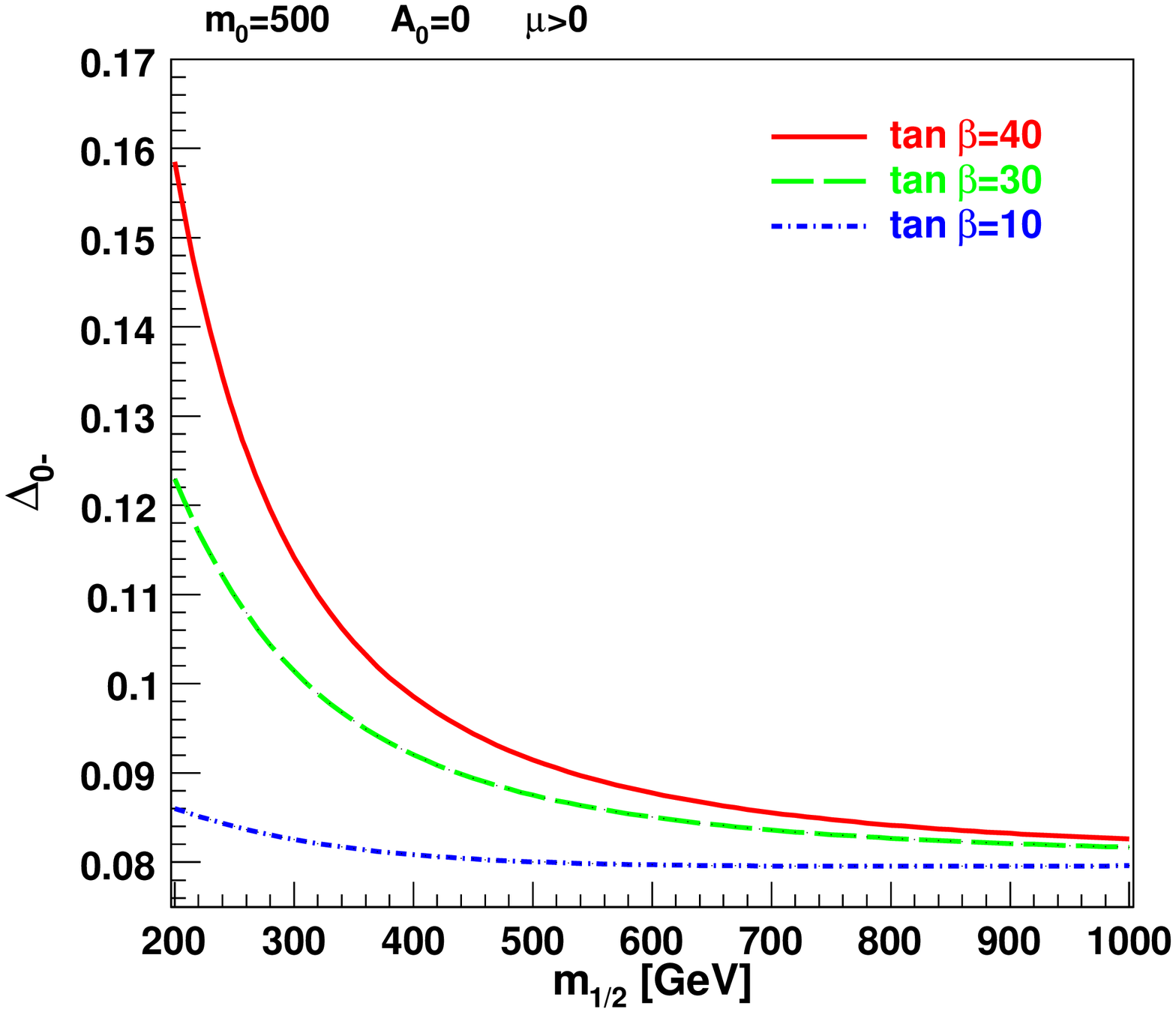}~\includegraphics[width=9cm,height=8cm]{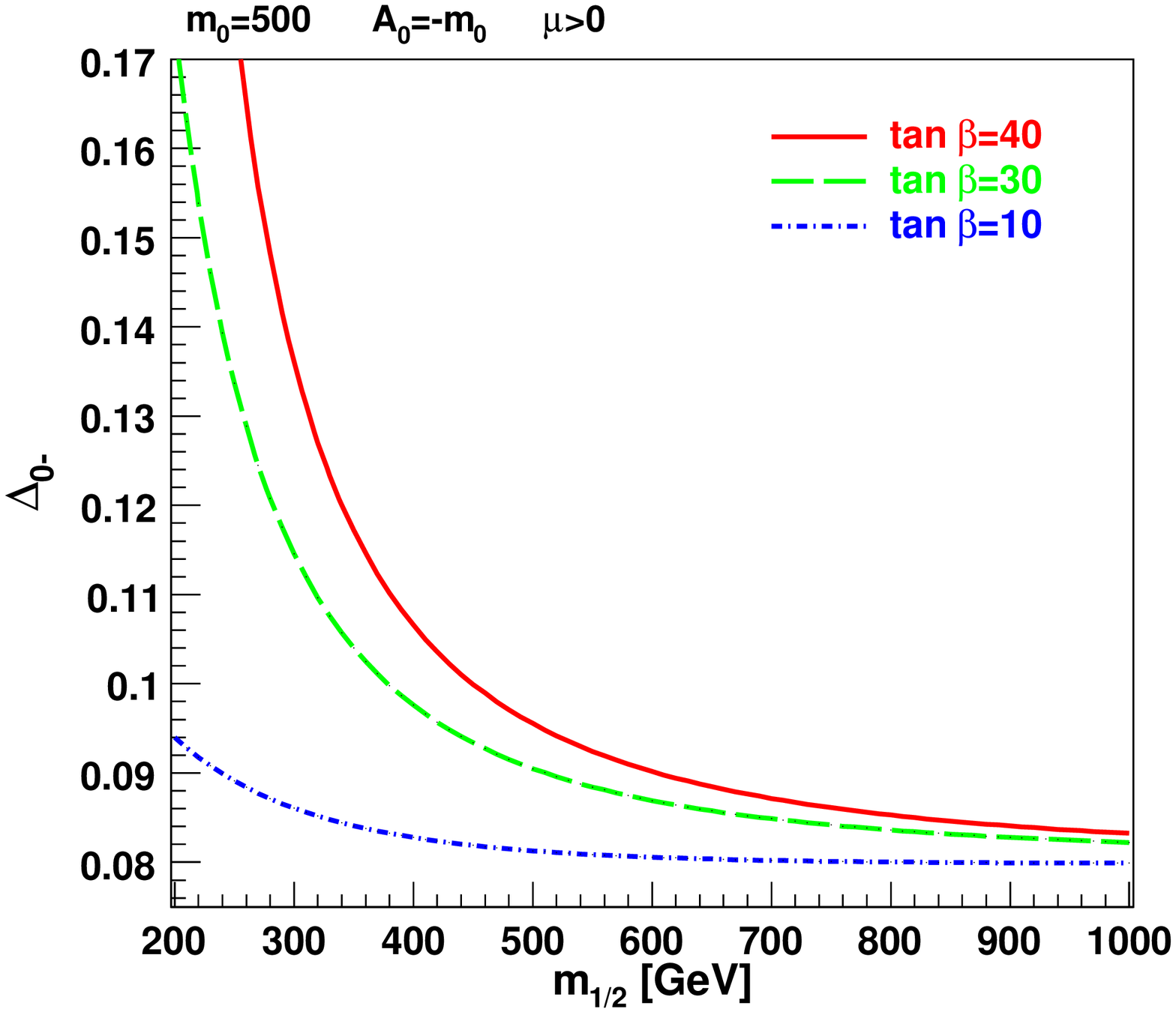}\\
\hspace*{-0.5cm}\includegraphics[width=8.78cm,height=8cm]{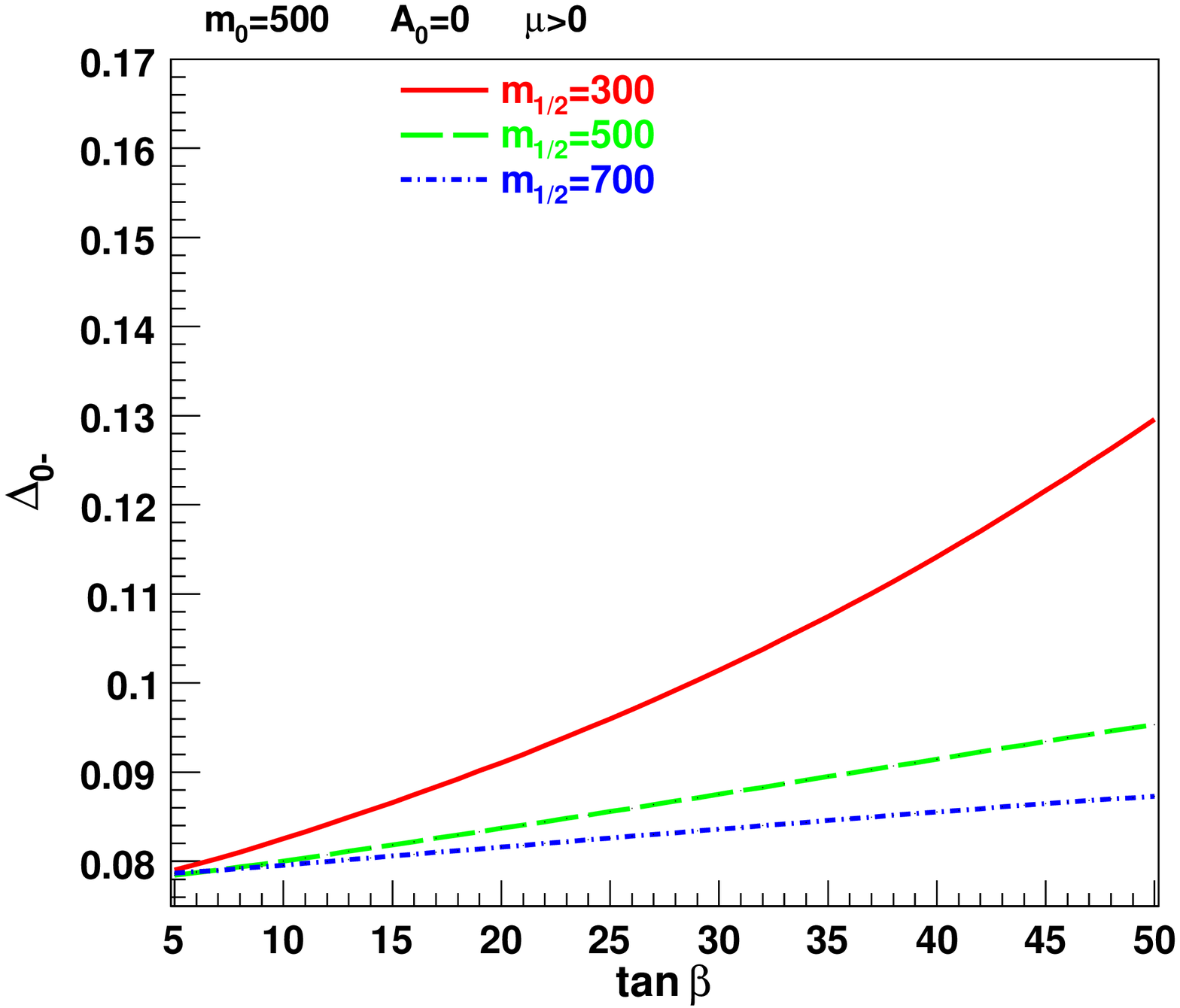}~~~~\includegraphics[width=8.78cm,height=8cm]{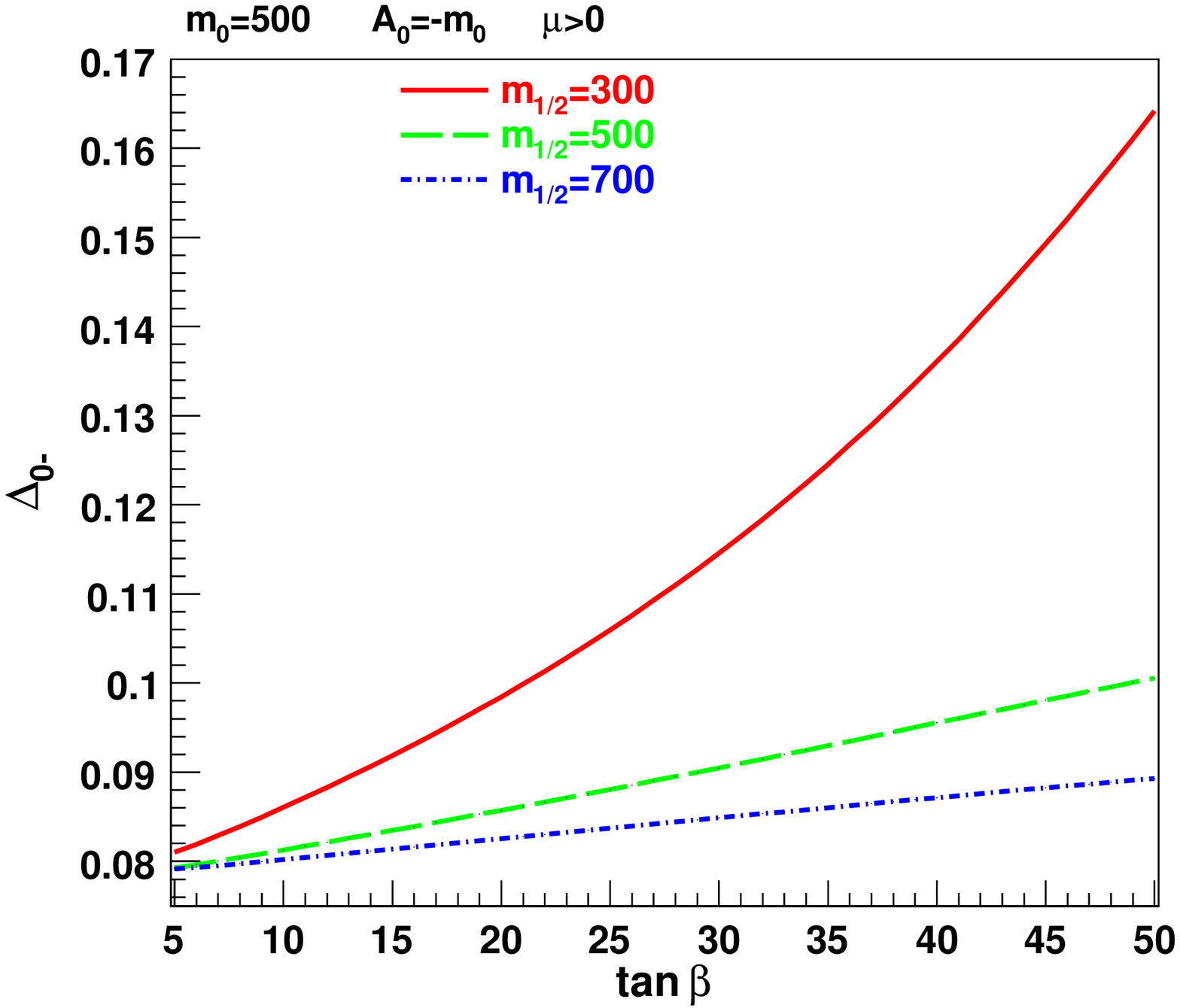}
\caption{Isospin asymmetry versus $m_{1/2}$ and $\tan\beta$ for $A_0=0$ and $A_0=-m_0$.}\label{2D}
\end{figure}%
\noindent We perform scans of the mSUGRA parameter space such that $m_0 \in [0,1000]$, $m_{1/2} \in [0,1000]$, $\tan\beta \in [0,50]$, $A_0 \in [-1000,1000]$ and for both signs of $\mu$. For $\mu < 0$, the supersymmetric contributions to the Wilson coefficients have the same sign as in SM. In this case, the latest experimental results are not sufficient to provide constraints on the mSUGRA parameter space. Moreover, $\mu < 0$ is disfavored by the $(g_{\mu} -2)$ measurements. For $\mu > 0$, the supersymmetric contributions to the Wilson coefficients can have a flipped-sign in comparison to the SM results, leading to a larger isospin breaking, and consequently, the experimental data can impose stringent constraints on the mSUGRA parameter space. Therefore, in the following we will only present results with $\mu>0$.\\
\\
An investigation of the $(m_{1/2},m_0)$ plane for $A_0=0$ is presented in Figure~\ref{A00}. In this figure, the area marked ``Isospin'' corresponds to the region excluded by the isospin breaking constraints, whereas the area marked ``$B\to X_s\gamma$'' corresponds to the region excluded by the inclusive branching ratio constraints. The ``Excluded'' area corresponds to the case where at least one of the particle masses does not satisfy the constraints of Table \ref{bounds}. And finally, ``Charged LSP'' is the cosmologically disfavored region when R-parity is conserved. The various colors represent the changing magnitude of the isospin asymmetry.\\
\\
First, we note that the isospin breaking for a set value of $(m_{1/2},m_0)$ increases with $\tan\beta$. Moreover, for a fixed $\tan\beta$, the asymmetry decreases with larger $m_0$ and $m_{1/2}$. Second, it should be pointed out that the constraints from isospin asymmetry are more stringent than the ones from inclusive branching ratio. However, for low $\tan\beta$ (like $\tan\beta=10$), constraints from both isospin asymmetry and branching ratio are not as restrictive as they are for larger values of $\tan\beta$.\\
\\
Figure~\ref{A0mm0} corresponds to the $(m_{1/2},m_0)$ plane for $A_0=-m_0$. %
The conventions are the same as in the previous figure. Again, we note that the isospin asymmetry is more sensitive to the model parameters than the inclusive branching ratio. 
A comparison between Figs. \ref{A00} and \ref{A0mm0} reveals that the isospin symmetry breaking is enhanced by a negative value of $A_0$ so that even for $\tan\beta$ as low as 10 it can produce appreciable constraints. Nonetheless, the global shapes remain similar.\\
\\
The effect of $\tan\beta$ on the isospin asymmetry is illustrated in Figure~\ref{tanb}. %
Indeed, the supersymmetric loop corrections which are proportional to the gluino mass and $\tan\beta$ can be quite large at high $\tan\beta$ limit. This arises from the Hall-Rattazzi-Sarid effect \cite{HRS}, and also from the top-quark Yukawa coupling \cite{D’Ambrosio}.
The enhancement of the isospin breaking by $\tan\beta$, particularly for smaller values of $m_{1/2}$, is clearly depicted in these graphs. These plots illustrate how stringent the isospin asymmetry bounds are at high $\tan\beta$, and also reveal the boost of the isospin breaking by a negative value of $A_0$.
In fact, the same trend is reported in $B_s \to \mu^+ \mu^-$ decay mode where the branching ratio can improve by as much as two orders of magnitude for large values of $\tan\beta$ \cite{choudhury,babu}.\\
\\
Figure~\ref{2D} illustrates the sensitivity of the isospin violation to $\tan\beta$ (and $m_{1/2}$) from a different perspective (and somewhat more clear as they are explicit non-contour plots) where the plots are done with two $A_0$ values: 0 and $-m_0$.\\
\\
To conclude this section, we have shown that the isospin asymmetry can provide stringent constraints on the mSUGRA parameter space, and appears to be even more contraining than the inclusive branching ratio.\\
\section{Summary}
\noindent In this article, we investigated the possibility for the isospin asymmetry in $B\to K^*\gamma$ decay mode to be an interesting observable to derive constraints on the supersymmetric parameter space. To obtain our results, we calculated the NLO supersymmetric contributions to the isospin asymmetry, using the effective Hamiltonian approach within the QCD factorization method and considering the minimal flavor violation. The mSUGRA parameter space was scanned, and the resulting isospin asymmetry for each point was compared to the experimental data from Babar and Belle. Our main conclusion of this comparison is that, provided $\mu>0$, the isospin asymmetry appears to be a powerful observable to constrain the mSUGRA parameter space producing even more stringent restrictions than the inclusive branching ratio. Among the different parameters, the values of $m_{1/2}$ and $\tan\beta$ appear to be restricted more significantly by the isospin symmetry breaking constraints.\\
\\
In this work we considered the mSUGRA model which has the advantage of having a fewer number of free parameters. However, mSUGRA's assumptions are in fact very strong and therefore extending this study to other SUSY models can be of interest.\\
\\
To conclude, with more accurate experimental data, we can hope the isospin asymmetry could reveal to be a very valuable observable to explore the supersymmetric parameter space.
\subsection*{Acknowledgments}
\noindent M.A.'s research is partially funded by a discovery grant from NSERC.  F.M. acknowledges the support of the McCain Fellowship at Mount Allison University.
\appendix
%
\vspace{1.3cm}
\section*{Appendix A: Wilson coefficients at $\mu_W$}\label{MW}
\renewcommand{\theequation}{A\arabic{equation}}%
\setcounter{equation}{0}%
\noindent The Wilson coefficients follow a perturbative expansion:
\begin{equation}
C_i(\mu_W) = C^{(0)}_i(\mu_W) + \frac{\alpha_s(\mu_W)}{4 \pi} C^{(1)}_i(\mu_W) + \cdots \;\;,
\end{equation}
where the $\alpha_s$ evolution writes \cite{PDG2006}:
\begin{equation}
\mbox{\small$\alpha_s(\mu)=\displaystyle\frac{4\pi}{\beta_0\ln(\mu^2/\Lambda^{(n_f)2})}\left[1-\frac{\beta_1}{\beta_0^2}\frac{\ln[\ln(\mu^2/\Lambda^{(n_f)2})]}{\ln(\mu^2/\Lambda^{(n_f)2})}+\frac{\beta_1^2}{\beta_0^4\ln^2(\mu^2/\Lambda^{(n_f)2})}\left(\left(\ln[\ln(\mu^2/\Lambda^{(n_f)2})]-\frac{1}{2}\right)^2+\frac{\beta_2\beta_0}{2 \beta_1^2}-\frac{5}{4}\right)\right]$}
\end{equation}\\
with
\begin{equation}
\beta_0=11-\frac{2}{3}n_f\;\;,\;\;\beta_1=102-\frac{38}{3}n_f\;\;\mbox{ and }\;\;
\beta_2=2857-\frac{5033}{9}n_f+\frac{325}{27}n_f^2\;\;,\label{beta01}
\end{equation}
$n_f$ being the number of flavors. $\Lambda^{(n_f)}$ is a dimensional parameter depending on the number of flavors. The numerical values of $\Lambda$'s in table \ref{tab:param} are based on the input $\alpha_s(M_Z)=0.1172$.\\
\\
The main contributions to the Wilson coefficients are classified into three categories: 1) those from the Standard Model, 2) charged Higgs contributions and 3) chargino contributions. The details of each contributing term are given in the following.
\vspace{0.3cm}
\subsection*{A.1~~~~Standard Model contributions}
\noindent The Standard Model contributions to the Wilson coefficients are adopted from Ref. \cite{ciuchini}. At leading order (LO), they read:
\begin{equation}
C_i^{SM(0)}(\mu_W) = C_i^{SM(0)}(\mu_W) = \left\{ \begin{array}{ccc}
0 &$\mbox{ for}$ & \mbox{ $i = 1,3,4,5,6$} \\
1 &$\mbox{ for}$ & \mbox{ $i = 2$} \\
\vspace{0.2cm} 
F_i^{(1)}(x_{tW}) & $\mbox{ for}$ & \mbox{ $i = 7,8$}\;\;, \\
\end{array} \right.
\end{equation}
where
\begin{equation}
x_{tW}=\displaystyle\frac{\bar m_t^2(\mu_W )}{M_W^2}\;\;,                                                     \end{equation} 
\begin{eqnarray}
F_7^{(1)}(x)&=&\frac{x(7-5x-8x^2)}{24(x-1)^3}+\frac{x^2(3x-2)}{4(x-1)^4}\ln x \;\;,\nonumber\\ F_8^{(1)}(x)&=&\frac{x(2+5x-x^2)}{8(x-1)^3}-\frac{3x^2}{4(x-1)^4}\ln x\;\;. \label{F781}
\end{eqnarray}
The NLO top quark running mass at a scale $\mu$ is given by \cite{PDG2006,chetyrkin}:
\begin{equation}
\bar m_t (\mu )=\bar m_t (m_t) \left[ \frac{\alpha_s (\mu)}{\alpha_s (m_t)}\right]^{\frac{\gamma_0^m}{2\beta_0}}
\left[ 1+ \frac{\alpha_s (m_t)}{4\pi}\frac{\gamma_0^m}{2\beta_0}\left(\frac{\gamma_1^m}{\gamma_0^m}-\frac{\beta_1}{\beta_0}\right)
\left( \frac{\alpha_s (\mu )}{\alpha_s (m_t)}-1\right) \right]\;\;,
\end{equation}
and
\begin{equation}
\bar m_t(m_t)=m_t\left[ 1-\frac{4}{3}\frac{\alpha_s (m_t)}{\pi}\right] \;\;,
\end{equation}
Here $m_t$ is the pole mass of the top quark. $\displaystyle\beta_0$ and $\displaystyle \beta_1$ are defined in eq.(\ref{beta01}) and:
\begin{equation}
\gamma_0^m= 8\;\;,\;\;\gamma_1^m=\frac{404}{3}-\frac{40}{9}n_f\;\;.
\end{equation}
The NLO corrections can be written as \cite{ciuchini}:
\begin{equation}
C_i^{SM(1)}(\mu_W)  = \left\{ \begin{array}{ccc} 15 + 6 \ln\dfrac{\mu_W^2}{M_W^2} & $\mbox{ for}$ & \mbox{  $i = 1$} \\
0 & $\mbox{ for}$ & \mbox{  $i = 2,3,5,6$} \\
E(x_{tW}) - \dfrac{2}{3} +\dfrac{2}{3}\ln\dfrac{\mu_W^2}{M_W^2} & $\mbox{ for}$ & \mbox{  $i = 4$}\\
G_i(x_{tW})+\Delta_i(x_{tW})\ln\dfrac{\mu_W^2}{M_W^2} & $\mbox{for}$ & \mbox{  $i = 7,8$}\;\;, 
\end{array} \right.
\end{equation}
where
\begin{eqnarray}
E(x) &=& \frac{x (-18 +11x + x^2)}{12 (x-1)^3} + \frac{x^2 (15 - 16 x + 4 x^2)}{6 (x-1)^4} \ln x-\frac{2}{3} \ln x\;\;,\\
G_7 (x) &=& \frac{- 436 + 2509 x - 10740 x^2  + 12205 x^3 + 1646 x^4}{486 (x-1)^4} + \frac{- 8 x + 80 x^2 -122 x^3 -16 x^4}{9 (x-1)^4} {\rm Li}_2 \left(1 - \frac{1}{x} \right) \nonumber \\ 
&& +\frac{208 - 1364 x + 3244 x^2 - 2262 x^3 - 588 x^4 -102 x^5} {81 (x-1)^5} \ln x +\frac{- 28 x^2 + 46 x^3 + 6 x^4}{3 (x-1)^5} \ln^2 x \;\;,~~~~~~~~~~\\
G_8(x) &=& \frac{- 508 + 610 x - 28209 x^2 -14102 x^3 + 737 x^4}{1296 (x-1)^4} + \frac{x + 41 x^2 + 40 x^3 - 4 x^4}{6 (x-1)^4} {\rm Li}_2 \left( 1 - \frac{1}{x} \right) \nonumber\\ 
&& +\frac{280 - 1994 x + 2857 x^2 + 4893 x^3 + 1086 x^4 -210 x^5} {216 (x-1)^5} \ln x +\frac{-31 x^2 - 17 x^3}{2 (x-1)^5} \ln^2 x \;\;,\\
\Delta_7(x) &=& \frac{208-1111x+1086x^2+383x^3+82x^4}{81(x-1)^4} +\frac{2x^2(14-23x-3x^2)}{3(x-1)^5}\ln x \;\;,\\
\Delta_8(x) &=& \frac{140-902x-1509x^2-398x^3+77x^4}{108(x-1)^4} +\frac{x^2(31+17x)}{2(x-1)^5}\ln x\;\;,
\end{eqnarray}
and where $\mbox{Li}_2$ is the usual dilogarithm function $\displaystyle\mbox{Li}_2(x)=-\int^x_0 dt\frac{\ln(1-t)}{t} \;$.
\vspace{0.3cm}
\subsection*{A.2~~~~Charged Higgs contributions}
\noindent At the LO, the relevant charged Higgs contributions to the Wilson coefficients are given by \cite{ciuchini}:
\begin{equation}
\delta C_{7,8}^{H(0)}(\mu_W) =\frac{A_u^2}{3} F_{7,8}^{(1)}(x_{tH^\pm})-A_uA_dF_{7,8}^{(2)}(x_{tH^{\pm}}) \;\;,
\end{equation}
with
\begin{equation}
A_u=-\frac{1}{A_d}=\frac{1}{\tan \beta}\;\;\;\mbox{ and }\;\;\; x_{tH^{\pm}}=\frac{\bar m_t^2(\mu_W )}{M_{H^\pm}^2}\;\;,
\end{equation}
where
\begin{eqnarray}
F_7^{(2)}(x)&=&\frac{x(3-5x)}{12(x-1)^2}+\frac{x(3x-2)}{6(x-1)^3}\ln x \;\;,\nonumber\\ F_8^{(2)}(x)&=&\frac{x(3-x)}{4(x-1)^2}-\frac{x}{2(x-1)^3}\ln x \;\;,
\end{eqnarray}
and $F_{7,8}^{(1)}$ are defined in eq.(\ref{F781}).\\
\\
At the NLO, the charged Higgs contributions are:
\begin{eqnarray}
\delta C_7^{(1)}(\mu_W) &=& G_7^H(x_{tH^{\pm}}) + \Delta_7^H(x_{tH^{\pm}}) \ln\frac{\mu_W^2}{M_H^2}-\frac49 E^H(x_{tH^{\pm}})\;\;,\\
\delta C_8^{(1)}(\mu_W) &=& G_8^H(x_{tH^{\pm}}) + \Delta_8^H(x_{tH^{\pm}}) \ln\frac{\mu_W^2}{M_H^2} -\frac16 E^H(x_{tH^{\pm}}) \;\;,
\end{eqnarray}
with
\begin{eqnarray}
G_7^H(x) &= &  A_dA_u\frac{4}{3}x\left[ \frac{4(-3+7x-2x^2)}{3(x-1)^3}{\rm Li}_2 \left( 1 - \frac{1}{x} \right)+\frac{8-14x-3x^2}{3(x-1)^4}\ln^2x +\frac{2(-3-x+12x^2-2x^3)}{3(x-1)^4}\ln x \right.\nonumber\\
&& \left. +\frac{7-13x+2x^2}{(x-1)^3}\right]+A^2_u\frac{2}{9}x\left[ \frac{x(18-37x+8x^2)}{(x-1)^4}{\rm Li}_2 \left( 1 - \frac{1}{x} \right)+\frac{x(-14+23x+3x^2)}{(x-1)^5}\ln^2x \right.\nonumber \\ 
&& \left .+\frac{-50+251x-174x^2-192x^3+21x^4}{9(x-1)^5}\ln x  +\frac{797-5436x+7569x^2-1202x^3}{108(x-1)^4}\right] \;\;,\\
\Delta_7^H(x) &= & A_dA_u\frac{2}{9}x\left[ \frac{21-47x+8x^2}{(x-1)^3}+\frac{2(-8+14x+3x^2)}{(x-1)^4}\ln x \right]\nonumber \\ 
&& +A_u^2\frac{2}{9}x\left[ \frac{-31-18x+135x^2-14x^3}{6(x-1)^4}+\frac{x(14-23x-3x^2)}{(x-1)^5}\ln x \right]\;\;,\\
G_8^H(x) &= & A_dA_u\frac{1}{3}x\left[ \frac{-36+25x-17x^2}{2(x-1)^3}{\rm Li}_2  \left( 1 - \frac{1}{x} \right)+\frac{19+17x}{(x-1)^4}\ln^2x +\frac{-3-187x+12x^2-14x^3}{4(x-1)^4}\ln x \right.\nonumber \\ 
&& \left. +\frac{3(143-44x+29x^2)}{8(x-1)^3}\right] +A^2_u\frac{1}{6}x\left[ \frac{x(30-17x+13x^2)}{(x-1)^4}{\rm Li}_2 \left( 1 - \frac{1}{x} \right)-\frac{x(31+17x)}{(x-1)^5}\ln^2x \right.\nonumber \\ 
&&  \left. +\frac{-226+817x+1353x^2+318x^3+42x^4}{36(x-1)^5}\ln x  +\frac{1130-18153x+7650x^2-4451x^3}{216(x-1)^4}\right]\;\;,
\end{eqnarray} 
\begin{eqnarray}
\Delta_8^H(x) &=& A_dA_u\frac{1}{3}x\left[ \frac{81-16x+7x^2}{2(x-1)^3}-\frac{19+17x}{(x-1)^4}\ln x \right]\nonumber \\
&& +A_u^2\frac{1}{6}x\left[ \frac{-38-261x+18x^2-7x^3}{6(x-1)^4}+\frac{x(31+17x)}{(x-1)^5}\ln x \right] \;\;,\\
E^H(x)&=&A_u^2\left[ \frac{x(16-29x+7x^2)}{36(x-1)^3}+\frac{x(3x-2)}{6(x-1)^4}\ln x \right] \;\;.
\end{eqnarray}
\vspace{0.3cm}
\subsection*{A.3~~~~Chargino contributions}
\noindent In the following, we use the notation $x_{ij}=\displaystyle\frac{m_i^2}{m_j^2}\;$. The masses of the sparticles are assumed such that for the squarks $m_{\tilde{q}_1} < m_{\tilde{q}_2}$, for the charginos $m_{\chi^{\pm}_1} < m_{\chi^{\pm}_2}$, and for the neutralinos $m_{\chi^0_1}<m_{\chi^0_2}<m_{\chi^0_3}<m_{\chi^0_4}$. The relevant chargino contributions read \cite{degrassi,gomez}:
\begin{eqnarray}
\delta C_{7,8}^{\chi}(\mu_s ) &=& - \sum_{k=1}^2 \sum_{i=1}^2 \left\{ \frac{2}{3} |\Gamma_{ki}|^2 \frac{M_W^2}{m_{\tilde t_k}^2} F_{7,8}^{(1)}(x_{\tilde t_k\chi_i^{\pm}}) + \Gamma_{ki}^*\Gamma_{ki}' \frac{M_W}{m_{\chi_i^{\pm}}} F_{7,8}^{(3)}(x_{\tilde t_k\chi_i^{\pm}})\right\} \nonumber\\
&& + \sum_{i=1}^2 \left\{ \frac{2}{3} |\tilde \Gamma_{1i}|^2\frac{M_W^2}{m_{\tilde{q}_{12}}^2} F_{7,8}^{(1)}(x_{\tilde{q}_{12}\chi_i^{\pm}})+ \tilde \Gamma_{1i}^*\tilde \Gamma_{1i}' \frac{M_W}{m_{\chi_i^{\pm}}} F_{7,8}^{(3)}(x_{\tilde{q}_{12}\chi_i^{\pm}})\right\}\;\;,
\end{eqnarray}
where $\mu_s$ is the SUSY scale, and $m_{\tilde{q}_{12}}$ is the common mass of the up and charm squarks, which we consider identical ($m_{\tilde{q}_{12}} \approx m_{\tilde{u}} \approx m_{\tilde{c}}$). Moreover, we have
\begin{eqnarray}
\Gamma_{ij} &=& D_{\tilde{t}1i}^{*}V^{*}_{j1} -\frac{\bar m_t(\mu_s)}{\sqrt 2 M_W\sin\beta} D_{\tilde{t}2i}^{*} V^*_{j2} \;\;,\nonumber\\
\Gamma_{ij}' &=& \frac{ D_{\tilde{t}1i}^{*} U_{j2}} {\sqrt 2 \cos\beta (1+\epsilon_{b}^*\tan\beta)} \;\;,
\end{eqnarray}
where $U$ and $V$ are the chargino mixing matrices, following the diagonalizing convention:
\begin{equation}
U \begin{pmatrix} M_2 & M_{W} \sqrt{2} \sin \beta \cr M_{W} \sqrt{2} \cos \beta & \mu \end{pmatrix} V^{-1}\;\;,
\end{equation}
and $D_{\tilde{q}}$ is the squark $\tilde{q}$ mixing matrix such as:
\begin{equation} 
D_{\tilde{q}} = \begin{pmatrix} \cos\theta_{\tilde{q}} & -\sin\theta_{\tilde{q}}  \cr \sin\theta_{\tilde{q}} & \cos\theta_{\tilde{q}} \end{pmatrix} \;\;,
\end{equation}
and $\epsilon_{b}$, which will be given below, is a two loop SUSY correction, whose effects are enhanced by factors of $\tan\beta$.\\
\\
$\tilde \Gamma_{ij}$ and $\tilde \Gamma_{ij}'$  are obtained  from $\Gamma_{ij}$ and $\Gamma_{ij}'$ by replacing the matrix $D_{\tilde{t}}$ by the unity matrix. The functions $F_{7,8}^{(3)}(x)$ are given by \cite{degrassi}:
\begin{eqnarray}
F_7^{(3)}(x) &=& \frac{(5-7x)}{6(x-1)^2} +\frac{x (3x-2)}{3(x-1)^3}\ln x \;\;,\nonumber\\
F_8^{(3)}(x) &=& \frac{(1+x)}{2(x-1)^2} -\frac{x}{(x-1)^3}\ln x \;\;.
\end{eqnarray}
The value of the chargino contributions at the scale $\mu_W$ is computed using:
\begin{eqnarray}
\delta C_7^\chi(\mu_W)&=& \eta_s^{-\frac{16}{3 \beta_0'}} \delta C_7^\chi(\mu_s)+\frac83 \left(\eta_s^{-\frac{14}{3 \beta_0'}}-\eta_s^{-\frac{16}{3 \beta_0'}}\right) \delta C_8^\chi(\mu_s)\;\;,\\
\delta C_8^\chi(\mu_W)&=& \eta_s^{-\frac{14}{3 \beta_0'}} \delta C_8^\chi(\mu_s)\;\;,
\end{eqnarray}
where $\eta_s \equiv \alpha_s(\mu_s)/\alpha_s(\mu_W)$ and $\beta_0' = -7$, which corresponds to six active flavors.\\
\\
In the following, we adopt the notations:
\begin{equation}
\cos \theta_{\tilde{q}} = D_{\tilde{q}11} \equiv  c_{\tilde{q}}\;\;,\;\;
\sin \theta_{\tilde{q}} = D_{\tilde{q}21} \equiv s_{\tilde{q}} \;\; .
\end{equation}
The leading $\tan\beta$ corrections are contained in the following formulas for $\epsilon_b$, $\epsilon_b^\prime$ and $\epsilon_t^\prime$, which are evaluated at scale $\mu_s$ \cite{micromegas,degrassi}:
\begin{eqnarray}
\epsilon_b  &=&\frac{2\,\alpha_s(\mu_s)}{3\,\pi} \frac{A_b/\tan\beta-\mu}{m_{\tilde{g}}} H(x_{\tilde{b}_1 \tilde{g}},x_{\tilde{b}_2 \tilde{g}})+\frac{ \tilde{y}_t^2(\mu_s)}{16\, \pi^2} \,\sum_{i=1,2} U_{i2}\frac{\mu/\tan\beta-A_t}{m_{\chi^\pm_i}}\,H(x_{\tilde{t}_1 \chi^\pm_i},x_{\tilde{t}_2 \chi^\pm_i}) \,V_{i2} \label{epsb} \nonumber\\
&&+ \frac{\alpha(M_Z)\mu M_2}{4 \sin^2\theta_W \pi} \times\\
&&\left[\frac{c_{\tilde{t}}^2}{m_{\tilde{t}_1}^2}H\left(\displaystyle\frac
{M_2^2}{m_{\tilde{t}_1}^2}, \displaystyle\frac
{\mu^2}{m_{\tilde{t}_1}^2}\right)+\frac{s_{\tilde{t}}^2}{m_{\tilde{t}_2}^2}H\left(\displaystyle\frac
{M_2^2}{m_{\tilde{t}_2}^2}, \displaystyle\frac
{\mu^2}{m_{\tilde{t}_2}^2}\right) \right. + \left. \frac{c_{\tilde{b}}^2}{2m_{\tilde{b}_1}^2} H\left(\displaystyle\frac
{M_2^2}{m_{\tilde{b}_1}^2}, \displaystyle\frac
{\mu^2}{m_{\tilde{b}_1}^2}\right)+\frac{s_{\tilde{b}}^2}{2 m_{\tilde{b}_2}^2}H\left(\displaystyle\frac
{M_2^2}{m_{\tilde{b}_2}^2}, \displaystyle\frac
{\mu^2}{m_{\tilde{b}_2}^2}\right)\right] \nonumber\;,
\end{eqnarray}\\
where $A_q$ is the trilinear coupling of the quark $q$. $y_q$ and $\tilde{y}_q$ are the ordinary and supersymmetric Yukawa couplings of the quark $q$ respectively. The function $H$ is defined by:
\begin{equation}
H(x,y) = \frac{x\ln\, x}{(1-x)(x-y)} + \frac{y\ln\, y}{(1-y)(y-x)}\;\;.
\end{equation}
Please note that we neglect the neutralino mixing matrices and we assume that the chargino masses are given by $\mu$ and $M_2$.\\
\begin{eqnarray}
\epsilon_b^\prime (t)&=& \frac{2\,\alpha_s(\mu_s)}{3\,\pi}\frac{A_b/\tan\beta-\mu}{m_{\tilde{g}}} \nonumber\times\\
&&\left[ c_{\tilde{t}}^2 c^2_{\tilde{b}}\,H(x_{\tilde{t}_1 \tilde{g}},x_{\tilde{b}_2 \tilde{g}}) + c_{\tilde{t}}^2 s^2_{\tilde{b}}
\,H(x_{\tilde{t}_1 \tilde{g}},x_{\tilde{b}_1 \tilde{g}}) + s_{\tilde{t}}^2 c^2_{\tilde{b}} \,H(x_{\tilde{t}_2 \tilde{g}},x_{\tilde{b}_2 \tilde{g}}) + s_{\tilde{t}}^2 s^2_{\tilde{b}} \,H(x_{\tilde{t}_2 \tilde{g}},x_{\tilde{b}_1 \tilde{g}})\right]  \nonumber \\
&& +\, \frac{ y_t^2(\mu_s)}{16\, \pi^2} \sum_{i=1}^{4}\,N_{i4}^*\frac{A_t-\mu/\tan\beta}{m_{\chi^0_i}} \times\\
&&\left[ c_{\tilde{t}}^2 c^2_{\tilde{b}}\,H(x_{\tilde{t}_2 \chi^0_i},x_{\tilde{b}_1 \chi^0_i}) + c_{\tilde{t}}^2 s^2_{\tilde{b}}\,    H(x_{\tilde{t}_2 \chi^0_i},x_{\tilde{b}_2 \chi^0_i})\label{epsbt} +\,s_{\tilde{t}}^2 c^2_{\tilde{b}}\, H(x_{\tilde{t}_1 \chi^0_i},x_{\tilde{b}_1 \chi^0_i}) + s_{\tilde{t}}^2 s^2_{\tilde{b}}\, H(x_{\tilde{t}_1 \chi^0_i},x_{\tilde{b}_2 \chi^0_i}) \right]\, N_{i3} \nonumber \\
&&+ \frac{\alpha(M_Z)\mu M_2}{4 \sin^2\theta_W \pi}  \nonumber\times\\
&& \left[\frac{c_{\tilde{b}}^2}{m_{\tilde{b}_1}^2}H\left(\displaystyle\frac
{M_2^2}{m_{\tilde{b}_1}^2}, \displaystyle\frac
{\mu^2}{m_{\tilde{b}_1}^2}\right)+\frac{s_{\tilde{b}}^2}{m_{\tilde{b}_2}^2}
H\left(\displaystyle\frac
{M_2^2}{m_{\tilde{b}_2}^2}, \displaystyle\frac
{\mu^2}{m_{\tilde{b}_2}^2}\right) \right.+ \left.\frac{c_{\tilde{t}}^2}{2 m_{\tilde{t}_1}^2}H\left(\displaystyle\frac
{M_2^2}{m_{\tilde{t}_1}^2}, \displaystyle\frac
{\mu^2}{m_{\tilde{t}_1}^2}\right)+\frac{s_{\tilde{t}}^2}{2m_{\tilde{t}_2}^2} H\left(\displaystyle\frac
{M_2^2}{m_{\tilde{t}_2}^2}, \displaystyle\frac
{\mu^2}{m_{\tilde{t}_2}^2}\right) \right] \nonumber\;.
\end{eqnarray}\\
$N$, in the above formula, represents the neutralino mixing matrix. The last correction reads:
\begin{eqnarray}
\epsilon_t^\prime (s)&=& -\frac{2\,\alpha_s}{3\,\pi} \frac{\mu+A_t/\tan\beta}{m_{\tilde{g}}} \left[ c_{\tilde{t}}^2
\,H(x_{\tilde{t}_2 \tilde{g}},x_{\tilde{s} \tilde{g}}) \;+\; s_{\tilde{t}}^2 H(x_{\tilde{t}_1 \tilde{g}},x_{\tilde{s} \tilde{g}}) \right]\\
&& +\, \frac{ y_b^2(\mu_s)}{16\, \pi^2} \sum_{i=1}^{4}\,N_{i4}^*\frac{\mu/\tan\beta}{m_{\chi^0_i}}
\, \left[ c_{\tilde{t}}^2 c^2_{\tilde{b}}\,H(x_{\tilde{t}_1 \chi^0_i},x_{\tilde{b}_2 \chi^0_i}) + c_{\tilde{t}}^2 s^2_{\tilde{b}}\,H(x_{\tilde{t}_1 \chi^0_i},x_{\tilde{b}_1 \chi^0_i}) \right.\nonumber \\
&&\left.+\,s_{\tilde{t}}^2 c^2_{\tilde{b}}\, H(x_{\tilde{t}_2 \chi^0_i},x_{\tilde{b}_2 \chi^0_i}) + s_{\tilde{t}}^2 s^2_{\tilde{b}}\, H(x_{\tilde{t}_2 \chi^0_i},x_{\tilde{b}_1 \chi^0_i}) \right]\, N_{i3}\nonumber\;\;.
\end{eqnarray}
The SM and charged Higgs contributions at the $\mu_W$ scale are affected by $\epsilon_b$, $\epsilon_b^\prime$ and $\epsilon_t^\prime$ as the following:
\begin{eqnarray}
\delta C_{7,8}^{(SM,\tan\beta)}(\mu_W)&=&\frac{\left[ \epsilon_b -\epsilon^\prime_b(t)\right] \tan\beta}{1+\epsilon_b\tan\beta} \, F_{7,8}^{(2)}(x_{tW})\;\;,\label{tbSM} \\
\delta C_{7,8}^{(H,\tan\beta)}(\mu_W)&=&-\frac{\left[ \epsilon^\prime_t(s) +\epsilon_b\right] \tan\beta}{1+\epsilon_b\tan\beta} \, F_{7,8}^{(2)}(x_{tH^\pm})\;\;.\label{tb2H}
\end{eqnarray}
Finally, the complete Wilson coefficients $C_{7,8}^{(0,1)}$ are found by adding the different contributions:
\begin{eqnarray}
C_{7,8}^{(0)}(\mu_W)&=&C_{7,8}^{SM(0)}(\mu_W)+\delta C_{7,8}^{H(0)}(\mu_W)+\delta C_{7,8}^{\chi}(\mu_W)+\delta C_{7,8}^{(SM,\tan\beta)}(\mu_W)+\delta C_{7,8}^{(H,\tan\beta)}(\mu_W)~~~~~~~~\\
C_{7,8}^{(1)}(\mu_W)&=&C_{7,8}^{SM(1)}(\mu_W)+\delta C_{7,8}^{H(1)}(\mu_W)\;\;.
\end{eqnarray}
\vspace{0.5cm}
\section*{Appendix B: Wilson coefficients at $\mu_b$}\label{mub}
\renewcommand{\theequation}{B\arabic{equation}}%
\setcounter{equation}{0}%
\noindent The Wilson coefficients at the lower scale $\mu_b = O(m_b)$ can be written as \cite{chetyrkin,buras}:
\begin{equation}
C_j(\mu_b)=C_j^{(0)}(\mu_b)+\frac{\alpha_s(\mu_b)}{4\pi}C_j^{(1)}(\mu_b) + \cdots \;\;,
\end{equation}
where, for $j=1\cdots6$:
\begin{equation}
C_j^{(0)}(\mu_b)=\sum_{i=3}^8 k_{ji}\eta^{a_i} \;\;\;,\;\;\;C_j^{(1)}(\mu_b)=\sum_{i=3}^8 \lbrack e_{ji}\eta E(x_{tW})+f_{ji}+g_{ji}\eta\rbrack \eta^{a_i}\;\;,
\end{equation}
with
\begin{equation}
\eta=\frac{\alpha_s(\mu_W)}{\alpha_s(\mu_b)} \;\;,
\end{equation}
\begin{eqnarray}
C_{7}^{(0)}(\mu_b) &=& \eta^\frac{16}{23} C_{7}^{(0)}(\mu_W) + \frac{8}{3}\left(\eta^\frac{14}{23} -\eta^\frac{16}{23}\right) C_{8}^{(0)}(\mu_W) + C_2^{(0)}(\mu_W)\sum_{i=1}^8 h_i \eta^{a_i}\;\;,\\
C_{8}^{(0)}(\mu_b) &=& \eta^\frac{14}{23} C_{8}^{(0)}(\mu_W) + C_2^{(0)}(\mu_W) \sum_{i=1}^8 \bar h_i \eta^{a_i}\;\;.
\end{eqnarray}
The next to leading coefficient $C^{(1)}_7$ is given by \cite{chetyrkin,buras}:
\begin{eqnarray}
C^{(1)}_7(\mu_b) &=& \eta^{\frac{39}{23}} C^{(1)}_7(\mu_W) + \frac{8}{3} \left( \eta^{\frac{37}{23}} - \eta^{\frac{39}{23}} \right) C^{(1)}_8(\mu_W) \nonumber \\ 
&& +\left( \frac{297664}{14283} \eta^{\frac{16}{23}}-\frac{7164416}{357075} \eta^{\frac{14}{23}} +\frac{256868}{14283} \eta^{\frac{37}{23}} -\frac{6698884}{357075} \eta^{\frac{39}{23}} \right) C_8^{(0)}(\mu_W) \\ 
&& +\frac{37208}{4761} \left( \eta^{\frac{39}{23}} - \eta^{\frac{16}{23}} \right) C_7^{(0)}(\mu_W) + \sum_{i=1}^8 (e_i \eta E(x_{tW}) + f_i + g_i \eta) \eta^{a_i} +\Delta C^{(1)}_7(\mu_b)\;\;,\nonumber
\end{eqnarray}
\def\thefootnote{\arabic{footnote}}%
\setcounter{footnote}{0}%
where in the $\overline{MS}$ scheme $\displaystyle \Delta C^{(1)}_7(\mu_b)=\sum_{i=1}^8 \left(\frac{2}{3}e_i  + 6 l_i \right) \eta^{a_i+1}\ln\frac{\mu_W^2}{M_W^2}\;$. The numbers $a_i$, $k_{ij}$, $e_{ij}$, $f_{ij}$, $g_{ij}$, $h_i$, $\bar h_i$, $e_i$, $g_i$ and $l_i$ are gathered in Table~\ref{tab:coef} \footnote{As we follow the operator basis from \cite{buras}, we also use the numerical values of this article.}. 
%
\begin{table}[!t]
\begin{center}
\begin{tabular}{|r|r|r|r|r|r|r|r|r|}
\hline
$i$ & 1 & 2 & 3 & 4 & 5 & 6 & 7 & 8 \\
\hline
$a_i $&$ \frac{14}{23} $&$ \frac{16}{23} $&$ \frac{6}{23} $&$-\frac{12}{23} $&$0.4086 $&$ -0.4230 $&$ -0.8994 $&$ 0.1456 $\\
\hline
$k_{1i} $& $0$ & $0$&$ \frac{1}{2} $&$ - \frac{1}{2} $&$0 $&$ 0 $&$ 0 $&$ 0 $\\
$e_{1i} $& $0$ & $0$&$ 0 $&$ 0 $&$0 $&$ 0 $&$ 0 $&$ 0 $\\
$f_{1i} $& $0$ & $0$&$ 0.8136 $&$ 0.7142 $&$0 $&$ 0 $&$ 0 $&$ 0 $\\
$g_{1i} $& $0$ & $0$&$ 1.0197 $&$ 2.9524 $&$0 $&$ 0 $&$ 0 $&$ 0 $\\
\hline
$k_{2i} $& $0$ & $0$& $ \frac{1}{2} $&$  \frac{1}{2} $&$0 $&$ 0 $&$ 0 $&$ 0 $\\
$e_{2i} $& $0$ & $0$&$ 0 $&$ 0 $&$0 $&$ 0 $&$ 0 $&$ 0 $\\
$f_{2i} $& $0$ & $0$&$ 0.8136 $&$ - 0.7142 $&$0 $&$ 0 $&$ 0 $&$ 0 $\\
$g_{2i} $& $0$ & $0$&$ 1.0197 $&$ - 2.9524 $&$0 $&$ 0 $&$ 0 $&$ 0 $\\
\hline
$k_{3i} $& $0$ & $0$& $ - \frac{1}{14} $&$ \frac{1}{6} $&$0.0510 $&$ - 0.1403 $&$ - 0.0113 $&$ 0.0054 $\\
$e_{3i} $& $0$ & $0$&$ 0 $&$ 0 $&$0.1494 $&$ -0.3726 $&$ 0.0738 $&$ -0.0173 $\\
$f_{3i} $& $0$ & $0$&$ -0.0766 $&$ - 0.1455 $&$-0.8848 $&$ 0.4137 $&$ -0.0114 $&$ 0.1722 $\\
$g_{3i} $& $0$ & $0$&$ -0.1457 $&$ - 0.9841 $&$0.2303 $&$ 1.4672 $&$ 0.0971 $&$ -0.0213 $\\
\hline
$k_{4i} $& $0$ & $0$& $ - \frac{1}{14} $&$ - \frac{1}{6} $&$0.0984 $&$ 0.1214 $&$ 0.0156 $&$ 0.0026 $\\
$e_{4i} $& $0$ & $0$&$ 0 $&$ 0 $&$0.2885 $&$ 0.3224 $&$ -0.1025 $&$ -0.0084 $\\
$f_{4i} $& $0$ & $0$&$ -0.2353 $&$ - 0.0397 $&$0.4920 $&$ -0.2758 $&$ 0.0019 $&$-0.1449 $\\
$g_{4i} $& $0$ & $0$&$ -0.1457 $&$ 0.9841 $&$0.4447 $&$ -1.2696 $&$ -0.1349 $&$ -0.0104 $\\
\hline
$k_{5i} $& $0$ & $0$& $ 0 $&$  0 $&$- 0.0397 $&$ 0.0117 $&$ - 0.0025 $&$ 0.0304 $\\
$e_{5i} $& $0$ & $0$&$ 0 $&$ 0 $&$-0.1163 $&$ 0.0310 $&$ 0.0162 $&$ -0.0975 $\\
$f_{5i} $& $0$ & $0$&$ 0.0397 $&$  0.0926 $&$0.7342 $&$ -0.1262 $&$ -0.1209 $&$ -0.1085 $\\
$g_{5i} $& $0$ & $0$&$ 0 $&$ 0 $&$-0.1792 $&$ -0.1221 $&$ 0.0213 $&$ -0.1197 $\\
\hline
$k_{6i} $& $0$ & $0$&$ 0 $&$  0 $&$0.0335 $&$ 0.0239 $&$ - 0.0462 $&$ -0.0112 $\\
$e_{6i} $& $0$ & $0$&$ 0 $&$ 0 $&$0.0982 $&$ 0.0634 $&$ 0.3026 $&$ 0.0358 $\\
$f_{6i} $& $0$ & $0$&$ -0.1191 $&$ - 0.2778 $&$-0.5544 $&$ 0.1915 $&$ -0.2744 $&$ 0.3568 $\\
$g_{6i} $& $0$ & $0$&$ 0 $&$ 0 $&$0.1513 $&$ -0.2497 $&$ 0.3983 $&$ 0.0440 $\\
\hline
$h_i $&$ 2.2996 $&$ - 1.0880 $&$ - \frac{3}{7} $&$ -\frac{1}{14} $&$ -0.6494 $&$ -0.0380 $&$ -0.0185 $&$ -0.0057 $\\
$\bar h_i $&$ 0.8623 $&$ 0 $&$ 0 $&$ 0 $&$ -0.9135 $&$ 0.0873 $&$ -0.0571 $&$ 0.0209 $\\
\hline
$e_i$ &$\frac{4661194}{816831}$&$ -\frac{8516}{2217}$ &$  0$ &$  0$ & $ -1.9043$  & $  -0.1008$ & $ 0.1216$  &$ 0.0183$\\
$f_i$ & $-17.3023$ & $8.5027 $ & $ 4.5508$  & $ 0.7519$& $  2.0040 $ & $  0.7476$  &$ -0.5385$  & $ 0.0914$\\
$g_i$ & $14.8088$ &  $ -10.8090$  &$ -0.8740$  & $ 0.4218$ & $  -2.9347$   & $ 0.3971$  & $ 0.1600$  & $ 0.0225$ \\
$l_i$ & $0.5784$ &  $ -0.3921$  &$ -0.1429$  & $ 0.0476$ & $  -0.1275$   & $ 0.0317$  & $ 0.0078$  & $ -0.0031$ \\
\hline
\end{tabular}
\end{center}
\caption{Useful numbers.\label{tab:coef}}
\end{table}
\vspace{0.9cm}
\section*{Appendix C: Isospin asymmetry}\label{isospin}
\renewcommand{\theequation}{C\arabic{equation}}%
\setcounter{equation}{0}%
\noindent To leading order the isospin asymmetry $\Delta_{0-}$ is given by \cite{kagan}:
\begin{equation}
\Delta_{0-} =\mbox{Re}(b_d-b_u) \;\;. \label{isospinC}
\end{equation} 
The spectator-dependent coefficients $b_q$ can be written as:
\begin{equation}
b_q = \frac{12\pi^2 f_B\,Q_q}{\bar m_b\,T_1^{B\to K^*} a_7^c}\left(\frac{f_{K^*}^\perp}{\bar m_b}\,K_1+ \frac{f_{K^*} m_{K^*}}{6\lambda_B m_B}\,K_{2q} \right)\;\;.
\end{equation}
The dimensionless functions $K_1$ and $K_{2q}$ are given by:
\begin{eqnarray}
K_1 &=& -\left( C_6(\mu_b) + \frac{C_5(\mu_b)}{N} \right) F_\perp\\
&&+\frac{C_F}{N}\,\frac{\alpha_s(\mu_b)}{4\pi}\,\left\{\left( \frac{\bar m_b}{m_B} \right)^2 C_8(\mu_b)\,X_\perp -C_2(\mu_b) \left[ \left(\frac43\ln\frac{m_b}{\mu_b} + \frac23 \right) F_\perp - G_\perp(x_{cb})\right] + r_1 \right\}\nonumber\;\;,
\end{eqnarray}
\begin{eqnarray}
K_{2q} &=& \frac{V_{us}^* V_{ub}}{V_{cs}^* V_{cb}}\left( C_2(\mu_b) + \frac{C_1(\mu_b)}{N} \right) \delta_{qu} + \left( C_4(\mu_b) + \frac{C_3(\mu_b)}{N} \right)\\
&&+\frac{C_F}{N}\,\frac{\alpha_s(\mu_b)}{4\pi} \left\{ C_2(\mu_b)\left( \frac43\ln\frac{m_b}{\mu_b} + \frac23 - H_\perp(x_{cb}) \right) + r_2 \right\} \;\;,\nonumber
\end{eqnarray}
where $N=3$ and $C_F=4/3$ are color factors, and
\begin{eqnarray}
r_1 &=& \left[\, \frac83\,C_3(\mu_b) + \frac43\,n_f \big(C_4(\mu_b)+C_6(\mu_b)\big) - 8\, \big(N C_6(\mu_b)+C_5(\mu_b)\big) \right] F_\perp \ln\frac{\mu_b}{\mu_0} + \dots \,, \nonumber\\
r_2 &=& \left[ -\frac{44}{3}\,C_3(\mu_b) - \frac43\,n_f \big(C_4(\mu_b)+C_6(\mu_b)\big) \right]\ln\frac{\mu_b}{\mu_0} + \dots \;\;, 
\end{eqnarray}
here $n_f=5$, and $\mu_0=O(m_b)$ is an arbitrary normalization scale. $r_1$ and $r_2$ are neglected in our calculations. The pole mass of the quarks can be deduced from the running quark mass at $\bar{m}_q$ \cite{PDG2006}:
\begin{eqnarray}
m_q&=&\bar{m}_q(\bar{m}_q)\left\{1+\frac{4 \alpha_s(\bar{m}_q)}{3\pi}+\left[-1.0414 \sum_k \left(1-\frac{4}{3}\frac{\bar{m}_{q_k}(\bar{m}_q)}{\bar{m}_q(\bar{m}_q)}\right)+13.4434\right]\left[\frac{ \alpha_s(\bar{m}_q)}{\pi} \right]^2\right.\\
&&\left. ~~~~~~~~~~~~~~~+\left[0.6527 n_{f_l}^2 - 26.655 n_{f_l} +190.595\right] \left[\frac{ \alpha_s(\bar{m}_q)}{\pi} \right]^3 \right\} \;\;,\nonumber
\end{eqnarray}
where the sum over $k$ extends over the $n_{f_l}$ flavors of the quarks $q_k$ lighter than the quark $q$. The functions $F_\perp$, $G_\perp(x_{cb})$, $H_\perp(x_{cb})$ and $X_\perp$ are convolution integrals of hard-scattering kernels with the meson distribution amplitudes, their values are given in Table \ref{tab:param}. The parameter $\displaystyle X=\ln(m_B/\Lambda_h)\,(1+\varrho\,e^{i\varphi})$ in this table parameterizes the logarithmically divergent integral $\int_0^1 dx/(1-x)$. We have evaluated the theoretical uncertainty by allowing $\varrho\le 1$ and the phase $\varphi$ to be arbitrary. $\Lambda_h \approx 0.5$ GeV is a typical hadronic scale.\\
\\
\def\thefootnote{\arabic{footnote}}%
\setcounter{footnote}{1}%
The coefficient $a_7^c$ is given by \cite{bosch}\footnote{A comparison has also been performed with the results of \cite{ali}. }:
\begin{table}[!ht]
\begin{center}
\begin{tabular*}{172mm}{@{\extracolsep\fill}|c|c|c|c|c|c|c|}
\hline
\multicolumn{7}{|c|}{CKM and coupling constant parameters}\\
\hline
$V_{us}$ & $V_{cb}$ & $\left|V_{ub}/V_{cb}\right|$ & $\mbox{Re}(V_{us}^* V_{ub}/V_{cs}^*V_{cb})$ & $\Lambda^{(4)}$ & $\Lambda^{(5)}$ & $\Lambda^{(6)}$\\
\hline
0.22 & $0.041 \pm 0.02$ & $0.085 \pm 0.025$ & $0.011\pm 0.005$ & 0.277 GeV & 0.200 GeV & 0.085 GeV\\
\hline
\end{tabular*}
\\
\vspace*{0.5cm}
\begin{tabular*}{172mm}{@{\extracolsep\fill}|c|c|c|c|c|}
\hline
\multicolumn{5}{|c|}{Parameters related to the $B$ mesons}\\
\hline
$m_B$ & $f_B$~ & $\lambda_B$~ & $H_2(x_{cb})$~ & $H_8$~\\
\hline
5.28 GeV & $200 \pm 20$ MeV & $(350 \pm 150)$ MeV & $(-0.27 \pm 0.06)+(-0.35 \pm 0.10)i$ & $0.70 \pm 0.07$ \\
\hline
\end{tabular*}
\\
\vspace*{0.5cm}
\begin{tabular*}{172mm}{@{\extracolsep\fill}|c|c|c|c|}
\hline
\multicolumn{4}{|c|}{Parameters related to the $K^*$ meson}\\
\hline
~~~~~ ~~~ $m_{K^*}$~~~ ~~~~ &  $f_{K^*}$~~~~~~ & $f_{K^*}^\perp$~~~~~~ & $T_1^{B\to K^*}$~~~~~~ \\
\hline
~~~~~ ~~~ 892 MeV~~~ ~~~~ & $226 \pm 28$ MeV~~~~~~  & $175 \pm 9$ MeV~~~~~~  & $0.30 \pm 0.05$~~~~~~ \\
\hline
\end{tabular*}
\\
\vspace*{0.5cm}
\begin{tabular*}{172mm}{@{\extracolsep\fill}|c|c|c|c|}
\hline
\multicolumn{4}{|c|}{Parameters related to the convolution integrals}\\
\hline
$F_\perp$ & $G_\perp(x_{cb})$ & $H_\perp(x_{cb})$ & $X_\perp$\\
\hline
\small$1.21\pm 0.06$ & \small$(2.82\pm 0.20)+(0.81\pm 0.23)i$ & \small$(2.32\pm 0.16)+(0.50\pm 0.18)i$ & \small$(3.44\pm 0.47)\,X-(3.91\pm 1.08)$ \\
\hline
\end{tabular*}
\\
\vspace*{0.5cm}
\begin{tabular*}{172mm}{@{\extracolsep\fill}|c|c|c|c|}
\hline
\multicolumn{4}{|c|}{Quark and W-boson masses}\\
\hline
$\bar m_b(\bar m_b)$ & $\bar m_c(\bar m_c)$~~~~ &  $m_t$~~~~~~ & $M_W$~~~~~~\\
\hline
~~~~~~~$4.20 \pm 0.07$ GeV~~~~~~~ & $1.25 \pm 0.09$ GeV~~~~~~ & $174.2 \pm 3.3$ GeV~~~~~~ & 80.4 GeV~~~~~~~\\
\hline
\end{tabular*}
\end{center}
\caption{The numerical values of the used parameters.\label{tab:param}}
\end{table}%
\begin{equation}
a^c_7(K^*\gamma) = C_7(\mu_b) + \frac{\alpha_s(\mu_b) C_F}{4\pi} \bigg( C_2(\mu_b) G_2(x_{cb})+ C_8(\mu_b) G_8\bigg) + \frac{\alpha_s(\mu_h) C_F}{4\pi} \bigg( C_2(\mu_h) H_2(x_{cb})+ C_8(\mu_h) H_8\bigg) \;\;,
\end{equation}
in which $\mu_h=\sqrt{\Lambda_h \mu_b}$, and
\begin{equation}
G_2(x_{cb}) = -\frac{104}{27}\ln\frac{\mu_b}{m_b}+ g_2(x_{cb}) \;\; , \;\;G_8 = \frac{8}{3}\ln\frac{\mu_b}{m_b} + g_8 \;\;,
\end{equation}
with
\begin{eqnarray}
g_2(x) &=& \frac{2}{9} \big( 48+30i\pi-5\pi^2-2i\pi^3 -36\zeta_3 +\left( 36+6i\pi-9\pi^2\right)\ln x+\left( 3+6i\pi\right) \ln^2\! x+\ln^3\! x \big)\, x \nonumber\\
&& {}+\frac{2}{9} \big( 18+2\pi^2 -2i\pi^3 +\left( 12-6\pi^2 \right)\ln x +6i\pi\ln^2\! x+\ln^3\! x\big) \,x^2 \\
&& {}+\frac{1}{27} \big( -9+112 i\pi-14\pi^2+\left(182-48i\pi\right)\ln x-126\ln^2\! x\big)\, x^3 -\frac{833}{162}-\frac{20i\pi}{27} +\frac{8\pi^2}{9} x^{3/2}\;\;,  \nonumber\\
g_8 &=& \frac{11}{3}-\frac{2\pi^2}{9}+\frac{2i\pi}{3} \;\;,
\end{eqnarray}
where $\zeta_3 \approx 1.2020569$ and $x_{cb}=\dfrac{\bar m^2_c}{\bar m^2_b}\;$. We have also:
\begin{equation}
H_2(x)=-\frac{2\pi^2}{3 N}\frac{f_B f^\perp_{K^*}}{T_1^{B\to K^*} m^2_B}\int^1_0 d\xi\frac{\Phi_{B1}(\xi)}{\xi}\int^1_0 dv\, h(1-v,x)\Phi_\perp(v) \;\;,
\end{equation}
where $h(u,s)$ is the hard-scattering function given by:
\begin{equation}
h(u,x)=\dfrac{4x}{u^2}\left[\mbox{Li}_2\!\left(\dfrac{2}{1-\sqrt{\dfrac{u-4x+i\varepsilon}{u}}}\right)+\mbox{Li}_2\!\left(\dfrac{2}{1+\sqrt{\dfrac{u-4x+i\varepsilon}{u}}}\right)\right]-\frac{2}{u} \;\;.
\end{equation}
$\Phi_\perp$ is the light-cone wave function with transverse polarization and $\Phi_{B1}$ is a distribution amplitude of the $B$ meson involved in the leading-twist projection. In a first approximation, $\Phi_\perp$ can be reduced to its asymptotic limit $\Phi_\perp(x)=6x(1-x)$ \cite{ball}. Finally, we can write:
\begin{equation}
H_8=\frac{4\pi^2}{3 N}\frac{f_B f^\perp_{K^*}}{T_1^{B\to K^*} m^2_B}\int^1_0 d\xi\frac{\Phi_{B1}(\xi)}{\xi}\int^1_0 dv\frac{\Phi_\perp(v)}{v} \;\;.
\end{equation}
The first negative moment of $\Phi_{B1}$ is parameterized by the quantity $\lambda_B$ such as $\displaystyle\int^1_0 d\xi\frac{\Phi_{B1}(\xi)}{\xi}=\frac{m_B}{\lambda_B}\;$.\\
The values of the different parameters can be found in Table~\ref{tab:param}.\\
\\
Using these relations altogether, it is then possible to calculate the isospin asymmetry from eq.(\ref{isospinC}).\\
\\
For the computation of the inclusive branching ratios, we also used the relations contained in the appendixes, together with those of Ref. \cite{kagan99}.\\
%

\end{document}